\documentclass[aip,reprint,floatfix,jcp]{revtex4-1}
\usepackage[]{graphicx}
\usepackage{color}
\usepackage{bm}

\newcommand{\colr}{}

\newcommand{\NA}{{N_{\rm A}}}
\newcommand{\eSi}{{^{28}{\rm Si}}}
\newcommand{\dd}{{d_{220}}}
\newcommand{\PTB}{{\rm PTB}}
\newcommand{\INRIM}{{\rm INRIM}}
\newcommand{\Tr}{{\rm Tr}}

\begin{document}

\title{A more accurate measurement of the $\bf \eSi$ lattice parameter}

\author{E.\ Massa}
\email[]{e.massa@inrim.it}
\author{C.\ P.\ Sasso}
\author{G.\ Mana}
\affiliation{INRIM -- Istituto Nazionale di Ricerca Metrologica, Str.\ delle Cacce 91, 10135 Torino, Italy}
\author{C.\ Palmisano}
\affiliation{UNITO -- Universit\`a di Torino, Dipartimento di Fisica, v. P.\ Giuria, 1 10125 Torino, Italy}


\date{\today}

\begin{abstract}
In 2011, a discrepancy between the values of the Planck constant measured by counting Si atoms and by comparing mechanical and electrical powers prompted a review, among others, of the measurement of the spacing of $\eSi$ \{220\} lattice planes, either to confirm the measured value and its uncertainty or to identify errors. This exercise confirmed the result of the previous measurement and yields the additional value $\dd=192014711.98(34)$ am having a reduced uncertainty.
\end{abstract}

\pacs{06.20.Jr, 06.20.F-, 06.30.Bp, 61.05.C-, 61.05.cp} 

\maketitle 


\section{Introduction}
Efforts are in progress on accurate determinations of the Planck, $h$, and Avogadro, $\NA$, constants.\cite{Bettin:2013} They are prompted by the proposal of a new kilogram definition based on a conventional value of the Planck constant, $\NA$ and $h$ being linked by the molar Planck constant, $\NA h$, which can be accurately measured.\cite{Mana:2012}

The most accurate way to determine $\NA$ is by counting the atoms in a single-crystal Si ball highly enriched with $\eSi$; a measurement was completed in 2011.\cite{Andreas:2011a,Andreas:2011b,Azuma:2015} The uncertainty associated with the determination of $h$ from this measurement is $3.0 \times 10^{-8} h$, {\colr but the measured value differed from the then most accurate result of the watt-balance comparisons of mechanical and electrical power.\cite{Steiner:2007}}

Although subsequent watt-balance determinations\cite{Sanchez:2014,Schlamminger:2014} gave $h$ values in substantial agreement with that obtained by atom counting, this prompted a reassessment of the $\NA$ uncertainty to identify whether errors were done. All the necessary measurements are being scrutinized and repeated aiming at a smaller uncertainty, thus carrying out stress tests of all the technologies to confirm that the intended performances are met. In the present paper we report about the measurement of the lattice parameter by means of combined x-ray and optical interferometry and give an additional result having a reduced uncertainty.

\section{$\NA$ measurement}
The $\NA$ value is obtained from measurements of the molar volume, $VM/m$, and lattice parameter, $a$, of a perfect and chemically pure silicon single-crystal. In a formula,
\begin{equation}\label{NA}
 \NA = \frac{8MV}{a^3 m} ,
\end{equation}
where $m$ and $V$ are the crystal mass and volume, $M$ is the mean molar mass, $a^3/8$ is the atom volume, and 8 is the number of atoms in the cubic unit cell. Since the binding energy of the Si atoms is about 5 eV and the mass of a Si atom is about 26 GeV, $M$ and $m$ can be viewed as the molar mass and mass of an ensemble of free atoms. To make the kilogram redefinition possible, the targeted accuracy of the measurement is $2\times 10^{-8} \NA$.

From (\ref{NA}), it follows that the $\NA$ determination requires the measurement of i) the lattice parameter -- by combined x-ray and optical interferometry\cite{Massa:2011a}, ii) the amount of substance fraction of the Si isotopes and, then, of the molar mass -- by absolute mass-spectrometry\cite{Pramann:2011,Narukawa:2014,Vocke:2014}, and iii) the mass and volume of nearly perfect crystal-ball having about 93 mm diameter.\cite{Picard:2011,Kuramoto:2011,Bartl:2011}

Silicon crystals may contain chemical impurities, interstitial atoms, and vacancies, which implies that the measured mass value does not correspond to that of an ideal Si crystal and that the crystal lattice may be distorted. This means that crystals must be characterized both structurally and chemically, so that the appropriate corrections are applied.\cite{Fujimoto:2011,Massa:2011b,Zakel:2011} The mass, thickness and chemical composition of the oxide layer covering the sphere must be taken into account; they are measured by optical and x-ray spectroscopy and reflectometry.\cite{Busch:2011}

\begin{figure}\centering
\includegraphics[width=8.4cm]{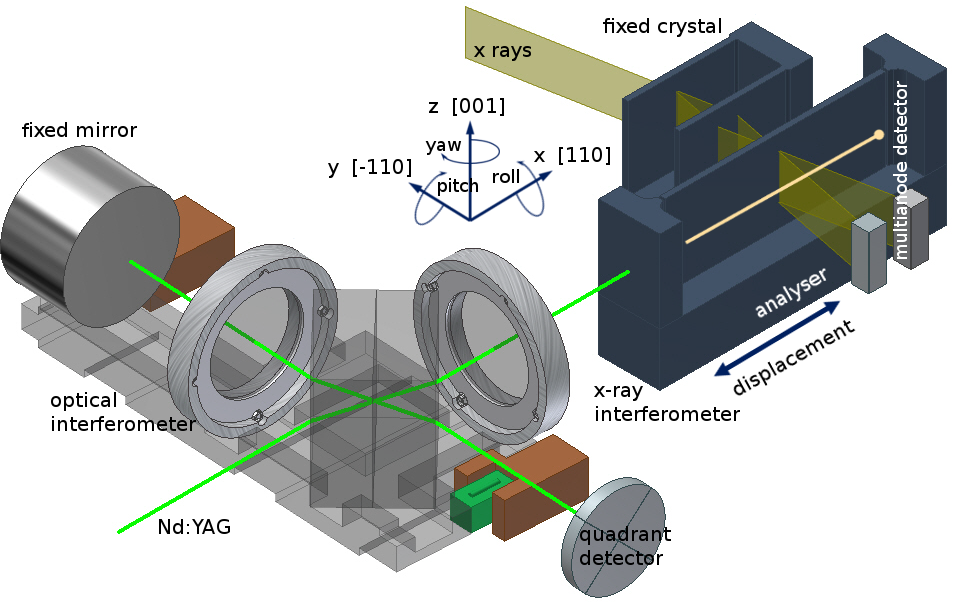}
\caption{Combined x-ray and optical interferometer. The yellow line indicates the continuation of the laser beam, at 21 mm from the analyser base, where the spacing of the diffracting planes was surveyed. The optical interferometer and fixed Si crystal rest on a common silicon plate (not shown). The analyser displacement and attitude (pitch and yaw angles) are optically sensed via quadrant detection of the interference pattern. The transverse, $y$ and $z$, displacements and roll rotation of the analyser are sensed via a reference 90$^\circ$ trihedron (resting on the same platform as the analyser) and three capacitive sensors faced to it (not shown).} \label{XROI}
\end{figure}

\section{Lattice parameter measurement}
\subsection{X-ray/optical interferometry}
The combined x-ray and optical interferometer used to measure the lattice parameter is described by Ferroglio {\it et. al}.\cite{Ferroglio:2008} As shown in Fig.\ \ref{XROI}, it consists of three blades, 1.20 mm thick, so cut that the \{220\} planes are orthogonal to the blade surfaces. X rays from a $(1\times 10)$ mm$^2$ Mo K$_{\alpha 1}$ line source are split by the first crystal and recombined, via a transmission crystal, by the third, called analyser. The interference pattern is imaged onto a multianode photomultiplier tube through a pile of eight NaI(Tl) scintillator crystals. The photomultiplier image projected on the analyser is $(1 \times 11.2)$ mm$^2$, with a pixel size of $(1 \times 1.4)$ mm$^2$.

When the analyser is moved along a direction orthogonal to the \{220\} planes, a periodic variation in the transmitted and diffracted x-ray intensities is observed, the period being the diffracting-plane spacing. The movement, up to 5 cm, requires control of the analyser attitude to within nanoradians and vibrations and position to within picometers. The analyser displacement and rotation are measured by optical interferometry; the necessary picometer and nanoradian resolutions are achieved by phase modulation, polarization encoding, and quadrant detection of the fringe phase. To eliminate the adverse influence of the refractive index of air and to ensure millikelvin temperature uniformity and stability, the apparatus is hosted  in a thermo-vacuum chamber.

We measured the lattice parameter of a number of natural Si crystals; the link between the results of these measurements and the measured value for the enriched crystal used to determine $\NA$ is given by Massa {\it et. al}.\cite{Massa:2011b} All the past measurements relied on the same optical interferometer, that served us since 1994. In order to exclude systematic effects, we assembled a new one and integrated it in the apparatus. The main novelties of the upgraded system are listed hereinbelow.

A 532 nm frequency-doubled Nd:YAG laser substituted for the previous 633 nm diode laser -- stabilised by frequency-offset technique against the frequency of an He-Ne laser which, in turn, was stabilized against component $a_{18}$ of the $^{127}$I$_2$ transition 11-5 R(127). The laser was better collimated, thus halving the correction for diffraction effects. The residual pressure in the vacuum chamber has been reduced to below 0.04 Pa. This makes any correction for the refractive index of the residual gas in the vacuum chamber inessential and ensures the calibration of the optical interferometer with a negligible uncertainty.

Contrary to our past measurement, the lattice spacing was surveyed along an horizontal line at 21 mm from the analyser base (instead of the previous 26 mm) and the correction for the self-weigh deformation was recalculated.

A new optical bench is clamped to the vacuum chamber; it collimates the laser beam, modulates the phase of the $\pi$-polarized component, and delivers it to the interferometer by a pointing mirror and a window of the vacuum chamber. The delivery, collimation, modulation, and pointing systems -- optical fiber, beam collimator and polarizer, phase modulator, and injection mirror -- have been rebuilt to conform to the new wavelength.

Previously, the orthogonality between the laser beam and the analyser was only occasionally checked. This was done by observing simultaneously, via a visual autocollimator placed -- when necessary -- outside the vacuum chamber, the analyser and laser beam through the output port of the interferometer. To gain the on-line control of the beam pointing, a home-made telescope picks up part of the beam delivered to the detector. In order to ensure stability, it is clamped on the same base plate as the x-ray/optical interferometer.

A plate beam-splitter was manufactured {\it ad-hoc} and substitutes for the cube beam-splitter previously used to ensure that the difference of the transmitted- and reflected-light paths is insensitive to the beam translations and rotations. Therefore, the components of the optical interferometer -- beam splitter, quarter-wave plates, and fixed mirror -- were replaced and assembled anew.

\begin{figure}\centering
\includegraphics[width=8.4cm]{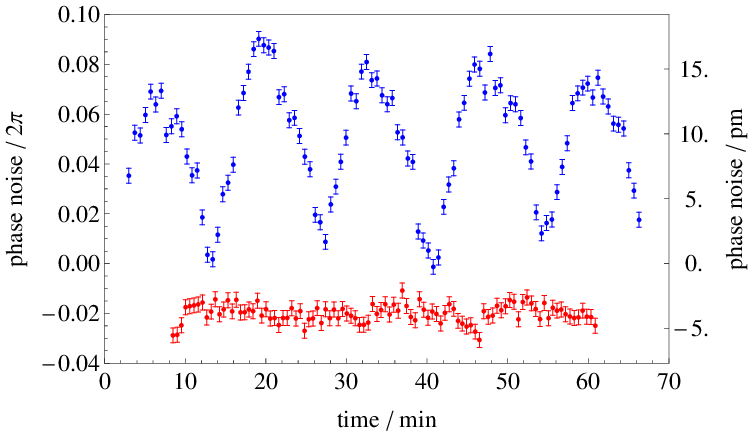}
\caption{Phase of the x-ray fringes when the analyser is locked to an integer optical interference-order. Top: noisy supply voltage of the optical-interferometer actuators. Bottom: upgraded voltage supply. The integration time of each measurement is 30 s. The right scale shows the corresponding analyser displacement.} \label{phase-noise}
\end{figure}

In order to make the interfering beams parallel, the components of the optical interferometer are cemented on a glass plate supported by three piezoelectric actuators. As shown in Fig. \ref{phase-noise}, a noise at a frequency of about 1 mHz caused phase instabilities between the x-ray and optical fringes. A new power supply was realized, having a sub part-per-million stability over the time scales, from 1 s to 1 h, relevant to the lattice parameter measurement. Eventually, the phase noise was reduced to the shot noise limit of the x-ray photon count.

The Physikalische Technische Bundesanstalt found a contamination of the surfaces of the x-ray interferometer by Cu, Fe, Zn, Pb, and Ca caused by the wet etching used by the INRIM to remove any residual stress due to surface damage after the crystal machining. The contamination was removed by cleaning the crystal in aqueous solutions of HF and (NH$_4$)$_2$S$_2$O$_8$.

The last upgrade concerned the temperature measurement. Since, the thermal expansion coefficient of $\eSi$ is about $2.6 \times 10^{-6}$ K$^{-1}$, the measured volumes of the Si balls and unit cell must refer to the same temperature to within a sub-millikelvin accuracy. Absolute temperature measurements are not necessary, but the temperature measuring-chains must be linked. We carried out more accurate and sensitive measurements of the Pt-thermometer resistance and linked our fixed point cells with those used to calibrate the measurements of the $\eSi$ ball temperature.

\subsection{Measurement procedure}
The measurement equation is
\begin{equation}
 a = \sqrt{8}\dd = \frac{\sqrt{8}m\lambda}{2n} ,
\end{equation}
where $\dd$ is the spacing of the \{220\} planes, $\sqrt{8}$ accounts for the different spacings of the \{100\} and \{220\} planes, and $n$ is the number of x-ray fringes in a step of $m$ optical fringes having period $\lambda/2$.

In practice, $\dd$ is determined by comparing the periods of the x-ray and optical fringes. This is done by measuring the x-ray fringe fraction at the ends of increasing steps $m\lambda /2$, where $m$ = 1, 10, 100, 1000, and 3570. We start from $\lambda /(2\dd) = n/m \approx 1385.95$ and measure the fringe fractions at the step ends with an accuracy sufficient for predicting the integer number of fringes in the next step. Consequently, $\lambda/(2\dd)$ is updated at each step. Eventually, measurements were carried out over 48 subsequent steps of $3570 \lambda /2$, 0.95 mm each for a total scanning length of 46 mm.

The least-squares method is applied to reconstruct the x-ray fringes and to determine their phases at the ends of each step.\cite{Mana:1991} Typical input data are 300 photon-counts over 100 ms time windows spaced by 4 pm; a typical sample contains six x-ray fringes, covers 1.2 nm, and lasts 30 s. Each $\dd$ measurement is the average of about nine values collected in measurement cycles where the analyser is repeatedly moved back and forth along the selected step. The visibility of the x-ray fringes approached 50\% with a mean brilliance of 500 counts s$^{-1}$ mm$^{-2}$. The crystal temperature is simultaneously measured with sub-millikelvin sensitivity and accuracy so that each $\dd$ value is extra\-po\-lated on-line to 20 $^\circ$C.

\subsection{Raw data}\label{data}
\begin{figure}\centering
\includegraphics[width=7.5cm]{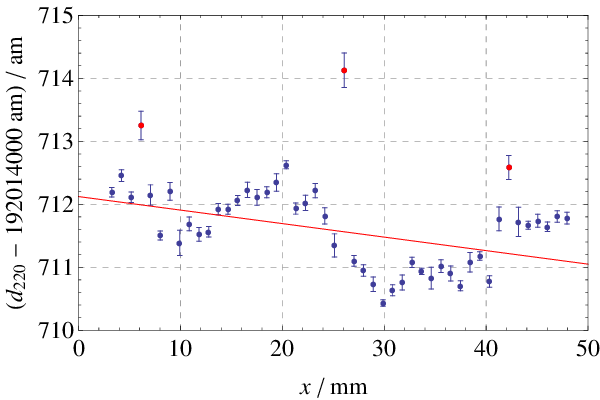}
\caption{Lattice spacing values measured on March 14 along the line shown in Fig.\ \ref{XROI}; x rays enter the analyser from the obverse face. All the values are extrapolated to 20 $^\circ$C, but not corrected for the systematic errors listed in Table \ref{table:1}. The bars indicate the uncertainty of the linear interpolation of the 8 detector-pixel values giving the value at 21 mm from the crystal base. The linear strain is due to the thermal gradient originated by the optical power injected into the crystal (from left side of the figure); the best fit line is also shown. The red dots indicate the outliers that were excluded from the subsequent analyses.} \label{d220a}
\end{figure}

Measurements were made over 48 subsequent analyser steps, 0.95 mm long. At each position, the lattice spacing was measured in the eight detector pixels and the 8 results were processed to obtain, by linear regressions, the lattice spacing values in 48 points of the horizontal line that is the continuation of the laser beam, at 21 mm from the analyser base (Fig.\ \ref{XROI}). A typical result is shown in Fig.\ \ref{d220a}.

\begin{figure}\centering
\includegraphics[width=7.5cm]{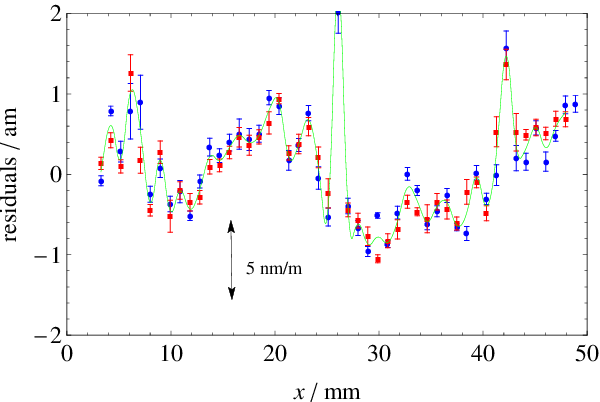}
\includegraphics[width=7.5cm]{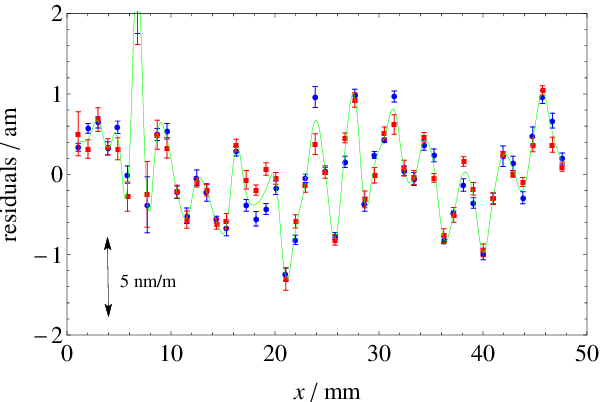}
\caption{Comparison of the variations of the measured lattice-spacing values along the line shown in Fig.\ \ref{XROI} and after the thermal strain has been removed. Top: obverse face of the analyser, February 02 (blue) and March 14 (red). Bottom: reverse face of the analyser, May 15 (blue) and June 05 (red).} \label{d220b}
\end{figure}

The figure shows a gradient of the lattice spacing; we discovered that it is correlated to the temperature gradient caused by the power, about 0.75 mW, injected by the laser beam in the analyser. The reduced thermal conductivity of the residual gas in the vacuum chamber -- because of the otherwise desirable low pressure -- contributed to worsening the problem. The way we coped with this problem is described in Sec.\ \ref{thermal-strain}.

Figure \ref{d220b} shows that, after the thermal strain is removed, the residuals and the outliers are repeatable from one measurement to the next -- also if carried out after one month. The head-on (obverse) and inverted (reverse) lay-outs correspond to the analyser crystal mounted as it was in the boule and in a reversed lay-out; after the reversal, the x rays cross the crystal in the opposite direction. The residuals and outliers repeatability, the scatter larger than the one expected by statistics, the different residual and outlier observed in the head-on and inverted surveys, and additional tests made by shifting the x-ray and optical baselines suggest that the outliers and residuals are caused by the analyser surfaces.

Apart from the outliers, that were again pinpointed, the profiles shown in Fig.\ \ref{d220b} are different from those given by Massa {\it et al.}.\cite{Massa:2011a} Although the support-point variability could partially explain the difference and in the past measurements the resolution and repeatability were not as good as today (we only spotted a correlation between the profiles taken with the same analyser orientation and no correlation between those taken with opposite orientations), we suspect that the difference is real and this substantiates our surface-effects allegation. A collaboration with the Leibniz Institutes f\"ur Oberfl\"achenmodifizierung in Liepzig is under way to develop machining technologies based on plasma etching and ion beams to gain a better control of the geometrical, physical, and chemical properties of the crystal surfaces.\cite{Paetzelt:2013,Paetzelt:2014}

\section{Analysis of the error budget}
\subsection{Statistics}
Each measurement is the mean, after eliminating the outliers (red in Fig.\ \ref{d220a}) and the thermal strain, of the survey results. The uncertainty of the mean is dominated by the variations of the measured values, that are supposed to be caused by local effects of the analyser surface. Therefore, when calculating the uncertainty, we took the residual correlation into account.

\subsection{Laser beam wavelength}
The frequency of the Nd:YAG line is locked to the component R(56) of the 32-0 transition of the $^{127}$I$_2$ molecule. It was measured to better than a $10^{-10}$ relative uncertainty; therefore, it does not contribute to the measurement uncertainty. To eliminate the influence of the refractive index of air, the experiment is carried out in vacuo. The residual pressure in the vacuum chamber has been reduced to less than 0.04 Pa. Since the air refractivity at the atmospheric pressure is $2.9\times 10^{-4}$, assuming that the pressure is in the interval from zero to 0.04 Pa with a uniform probability, the relevant correction is 0.058(33) nm/m.

\subsection{Laser beam diffraction}\label{diffraction-section}
The period of the interference fringes is not equal to the plane-wave wavelength. In the case of the interference of two identical paraxial beams -- whose angular spectra are strongly concentrated around a wave-vector having $\omega/c$ modulus, the difference between the period of the integrated interference pattern, $\lambda_e$, and the plane-wave wavelength, $\lambda=2\pi c/\omega$, is\cite{Mana:1999a,Mana:2006}
\begin{equation}\label{diffraction-correction}
 \frac{\lambda_e - \lambda}{\lambda} = \frac{(1-x_0^2/w_0^2)\Tr(\bm{\Gamma})}{2} + \frac{\alpha^2}{2} + \frac{\gamma^2}{2} ,
\end{equation}
where the optical-path difference of the interferometer arms is assumed to be much smaller than the Rayleigh length, $x_0$ is the offset between the beam axes measured at the beam waists, $w_0$ is the 1/e$^2$ spot radius at the beam waist, $\bm{\Gamma}$ is the central second-moment matrix of the angular power-spectrum of the beams, $2\alpha$ is the misalignment between the beam axes, and $\gamma$ is the beam deviation from a normal incidence on the analyser.

The $\Tr(\bm{\Gamma})$ term originates from diffraction and depends on the spread of the transverse impulse of the photons. It holds for any paraxial beam, no matter if its profile is Gaussian or not\cite{Mana:1999a}; actually, it is obtained under a coaxial-beam assumption, i.e., when $x_0/w_0=\alpha=0$, and subsequently generalized to non-coaxial beams, but under a Gaussian-beam assumption.\cite{Mana:2006} It must be noted that, in the case of Gaussian beams having cylindrical symmetry, $\Tr(\bm{\Gamma})/2=\theta_0^2/4$, where $\theta_0$ is the far-field divergence. We measured the angular power-spectrum of the beams emerging from the interferometer by using the Fourier transforming properties of a lens. Next, (\ref{diffraction-correction}) is calculated from the central second-moment matrix of the focal plane image, which is recorded by a videocamera.

\begin{figure}\centering
\includegraphics[width=7.5cm]{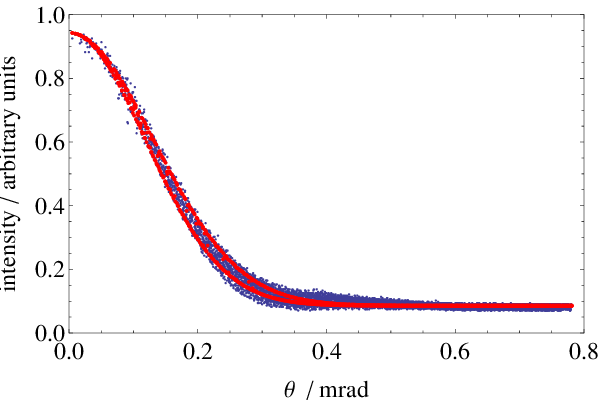}
\caption{Radial profile of the focal plane image of the laser beam. The solid lines are minimum and maximum values of the radial profile of the best bivariate Gaussian function fitting the data.} \label{FP-image}
\end{figure}

Owing to the 8 bit resolution and dark-noise of the camera that we are presently using, a calculation of $\bm{\Gamma}$ based on a discrete approximation of the relevant integrals is unreliable.\cite{Mana:2001} Therefore, it was estimated by fitting a bivariate Gaussian function to the focal-plane image; an example is shown in Fig.\ \ref{FP-image}. The uncertainty associated to the $\Tr(\bm{\Gamma})$ estimate is small, typically, less than 1\%.

\begin{figure}\centering
\includegraphics[width=7.8cm]{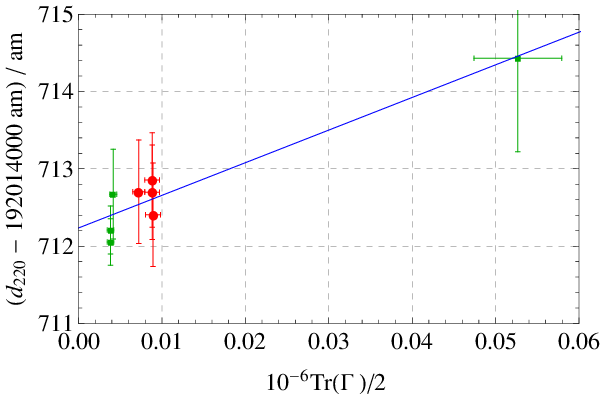}
\caption{Corrected $\dd$ values {\it vs.}~the applied correction for the diffraction of the laser beam. The measurements are made from 2010 to 2014 by using laser beams differently collimated. Bullets (red): He-Ne laser source @ 633 nm; squares (green): frequency doubled Nd:YAG laser source @ 532 nm. If the corrections are correctly estimated, the data are expected to lie on an horizontal line.} \label{diffraction}
\end{figure}

To check the correction estimate, we examined the results of a number of $\dd$ measurements carried out from 2010 to 2014 with different beams. The results shown in Fig.\ \ref{diffraction} suggest that we overestimate the correction. Subsequent investigations did not shed light on this problem, but a study of the interference of wavefronts differently perturbed in the separate arms of the interferometer -- where (\ref{diffraction-correction}) does not hold exactly -- seems to support an overestimation; more details will be given in a separate paper. Another hypothesis is a wrong estimate of the center of mass of the focal-plane image, which implies a correction always larger than true. In addition, since a single datum -- corresponding to the largest beam divergence, see Fig.\ \ref{diffraction} -- dictates the regression line, we may be mislead by a measurement error. Owing to the smallest beam divergence in the present set-up, the result we are reporting agrees with the value extrapolated to a zero correction from the data in Fig.\ \ref{diffraction-correction}. Therefore, we did not correct the $\Tr(\bm{\Gamma})/2 = 3.978\times 10^{-9}$ value; but, cautiously, increased its uncertainty to $15\%\Tr(\bm{\Gamma})/2$.

The interfering beams are kept parallel to within a $2\alpha = 1$ $\mu$rad maximum misalignment by levelling the phase in four quadrants of the interference pattern via the piezoelectric supports (pitch) and inertial drivers (yaw) of the interferometer base-plate. Consequently, the $\alpha^2/2$ term in (\ref{diffraction-correction}) is irrelevant and was omitted.

\subsection{Laser beam alignment}\label{alignment-section}
\begin{figure}\centering
\includegraphics[width=7.5cm]{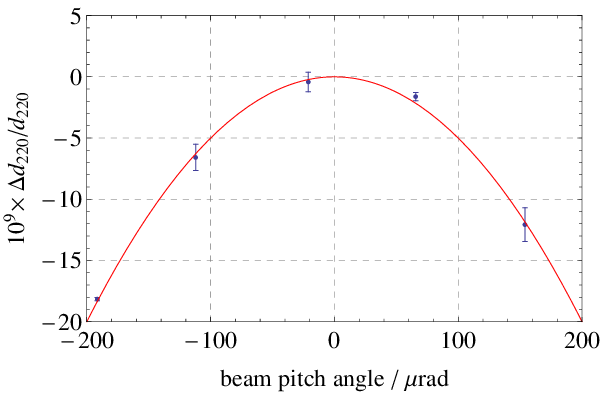}
\caption{Measured $\dd$ values {\it vs.} the pitch angle of the laser beam; the angle origin is set in the maximum of the parabola (red line) that best fit the data.} \label{allineamento}
\end{figure}

When assembling the apparatus, the laser beam deviation $\gamma$ from a normal incidence on the analyser was nullified with the aid of an autocollimator looking at both the analyser and beam from the interferometer output-port. Next, as shown in Fig.\ \ref{allineamento}, we carried out a number of $\dd$ measurements while $\gamma$ was purposely changed along two orthogonal directions and its variations were recorded by an on-line telescope. The telescope is mounted, inside the vacuum chamber, on the same base plate of the x-ray/optical interferometer and picks up part of the output beam. After two parabola were fitted to the measured $\dd$ values, the beam direction corresponding to the maxima -- hence, to a supposed normal incidence -- was identified to within $\mu$rad uncertainty and maintained in the telescope optics with an uncertainty of 20 $\mu$rad. Eventually, the laser beam was kept parallel to that direction and the telescope readings recorded for subsequent analyses.

After we completed the measurements and removed the interferometer from the apparatus, we realised that the operation of the optical interferometer may be liable to a systematic error. This problem will be examined in a separate paper, but, since it relates to the assessment of the measurement uncertainty, we outline it shortly.

In the case of a pointing error $\gamma$, an analyser displacement $s$ shears the interfering beams by $2s\gamma$. Hence, the measure beam goes through different parts of the optics crossed in its way to the detector. This shear changes the optical-path length by $(2\gamma s)\beta\Delta n$, where $\Delta n$ and $\beta$ are the refractivity and the relevant component of the vertical angle of a wedge that, in a simplified model, substitutes for the optics in the way from the analyser to the detector. In addition, because of the wavefront curvature, the beam shear is sensed by the interferometer as a rotation equal to
\begin{equation}\label{m2}
 \Omega = \frac{s\gamma}{R} ,
\end{equation}
where $1/R$ is the wavefront curvature. Therefore, the measured $\dd$ value is
\begin{equation}\label{parabola}
 d_m = \dd \left[ 1 - \frac{1}{2}\gamma^2 + \left( \beta\Delta n + \frac{b}{2R} \right)\gamma \right] ,
\end{equation}
where $b$ is the Abbe's offset between the laser and x-ray beam-centroids. According to (\ref{parabola}), $d_m$ is maximum when $\gamma=\beta\Delta n + b/(2R)$, not when $\gamma=0$ as assumed in the alignment procedure. It must be noted that, when $d_m$ is maximum, the sensed rotation is not zero -- as expected if $\gamma=0$ rad -- but,
\begin{equation}\label{m3}
 \Omega = \frac{s\beta\Delta n}{R} ,
\end{equation}
where, for the sake of simplicity, we assumed $b=0$ mm.

\begin{figure}\centering
\includegraphics[width=7.5cm]{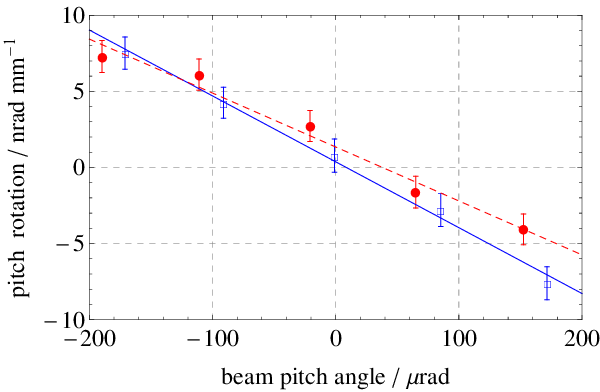}
\caption{Parasitic pitch of the analyser motion vs.\ the pitch angle of the laser beam. The analyser motion is servoed so as to nullify the differential signals detected by the optical interferometer. Measurements were done in February (squares, obverse face) and May (bullets, reverse face). The lines (solid and dashed, respectively) that best fit the data are also shown. The angle origin is set in the maximum of the parabola that best fit the measured $\dd$ values (see Fig.\ \ref{allineamento}).} \label{pitch-1}
\end{figure}

Since the analyser attitude is servoed so as to nullify signal of the angle interferometer -- the differential phase between the quadrants of the interference pattern, the shear is counteracted by an analyser rotation. Eventually, the pitch component of this rotation is disclosed by a $\dd$ gradient in the different detector pixels. The pitch explaining the gradient observed with a varying alignment of the laser beam is shown in Fig.\ \ref{pitch-1}. When the beam is aligned in such a way that the $\dd$ value is maximum, the pitch was equal to 0.4 nrad/mm in February and, after the analyser reversal and realignment, to 1.3 nrad/mm in May. The yaw rotation might be similar, but, presently, we cannot detect it.

\begin{figure}\centering
\includegraphics[width=7.5cm]{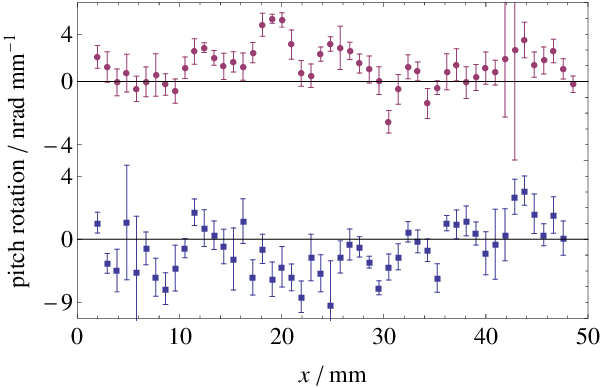}
\caption{Parasitic pitch of the analyser motion measured over 0.95 steps in 48 subsequent analyser positions. Surveys were carried out on February 02 (bottom, obverse face) and May 15 (top, reverse face). The laser beam was pointed so as to ensure that the measured $\dd$ value was maximum and the analyser was servoed so as to nullify the differential signals detected by the optical interferometer.} \label{pitch-2}
\end{figure}

Shear strains of the crystal lattice and surface effects (see Sec.\ \ref{data}) mimic the same gradients. Therefore, we cannot unambiguously explain a $\dd$ gradient by a parasitic pitch rotation. However, a number of additional tests excluded large strains. For instance, Fig.\ \ref{pitch-2} shows the parasitic pitch of the analyser that explains the $\dd$ gradients observed in the March 14 and May 15 surveys, where the laser beam was aligned in such a way that the $\dd$ measure is maximum. Since the only survey difference is the reversed analyser alignment, if we had observed a lattice strain, the two plots should be similar. Contrary, they are not; the mean pitch was 1.0 nrad/mm in March and $-0.5$ nrad/mm in May.

According to (\ref{m2}), the radius of curvature of the interfering wavefronts, $R=25(1)$ m, can be estimated from the mean slope, $s/R = 40(2)\times 10^{-6}$ mm$^{-1}$, of the lines that best fit the data in Fig.\ \ref{pitch-1}. By explaining the $\dd$ gradients in terms of parasitic pitch rotations and (\ref{m3}), we estimate that the pitch component of $\bm{\beta}$ is 50 $\mu$rad from the March data and $-25$ $\mu$rad from the May one. Since no change was made in the optical interferometer, these contradictory estimates exclude wedge angles much greater than, say, 50 $\mu$rad.

In order to estimate the needed correction and the associated uncertainty, we used $\Delta n \approx 1/2$ and calculated the mean and variance of the $c_\gamma = (\gamma-\beta)\gamma\dd/2$ correction, where $\beta=0(50)$ $\mu$rad and $\gamma = 0.5\beta \pm 20$  $\mu$rad are independently normally distributed. The final result is $c_\gamma = -0.11(48)$ nrad$^2$.

\subsection{Laser beam walk}
The beam walk refers to the transverse motion of the interfering beams through the optical components. It originates from different effects causing the beams to move across imperfect surfaces or wedged optics. The effect of walks caused by a tilted incidence on the analyser was investigated in Sec.\ \ref{alignment-section}. In our previous set-up, the beam-splitter imperfection, combined with tilts of the apparatus baseplate with respect to the laser beam, caused systematic differential variations of the optical paths through the interferometer that required {\it ad hoc} corrections. In the new apparatus, we made this problem harmless by using a plate beam-splitter, having a parallelism error less than 10 $\mu$rad, and by controlling electronically the baseplate level and tilt to within 25 nm and 70 nm/m, over any analyser (short or long) displacement. The differential beam walk due to analyser parasitic rotations is irrelevant because rotations are less than 1 nrad/mm (see Sec.\ \ref{alignment-section}) and the detector distance is less than 0.5 m.

{\colr The mechanical load driving the analyser carriage -- from the outside of the vacuum chamber -- causes the inside apparatus to sag and to yaw with respect to the laser beam. During the motion, the relevant beam walks are quite large, up to 1 $\mu$m and 5 $\mu$rad, but, after any displacement, the mechanical link between the apparatus and the driving system is removed, thus allowing the equilibrium position to be restored and, in principle, any beam walk to be nullified. It is difficult to estimate if, after averaging over the 48 displacements, there is a residual systematic walk of the interfering beams. We assumed that the mean walk over a 1 mm displacement is uniformly random in the $[-0.1,0.1]$ $\mu$m interval, a 10\% of the maximum observed without mechanical disconnection and averaging. We assumed also that the differential wedge-angle between the end surfaces of the separate paths through the interferometer is uniformly random in the $[-10,10]$ $\mu$rad interval. Consequently, the difference of the optical paths associated to the beam walk is zero, with standard deviation of 0.577 nm/m.}

\begin{figure}\centering
\includegraphics[width=7.5cm]{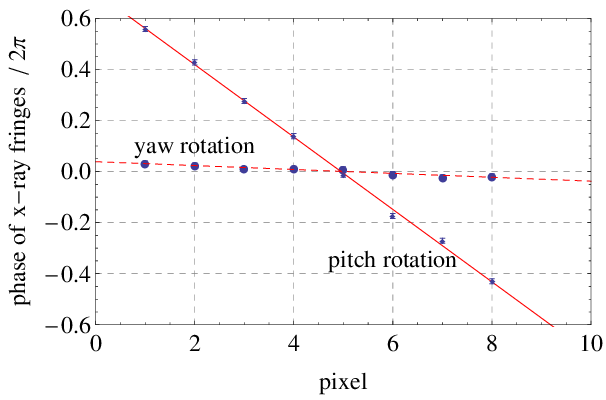}
\caption{Vertical variations of the x-ray fringe phase when the analyser is rotated -- while keeping the displacement measured by the optical interferometer null -- about the $y$ (solid line) and $z$ (dashed line) axes.} \label{Abbe}
\end{figure}

\begin{figure}\centering
\includegraphics[width=7.5cm]{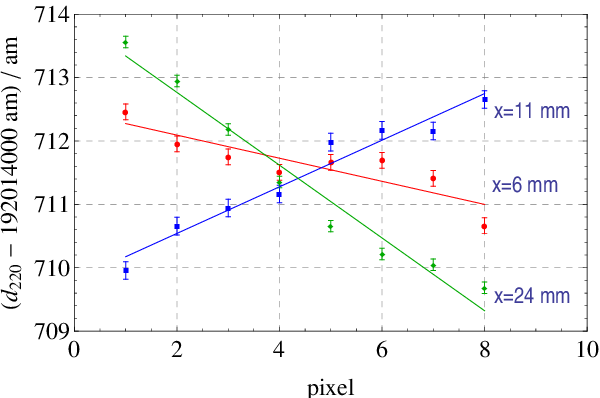}
\caption{Vertical variations of the $\dd$ value measured over different 0.95 mm steps. The coordinate of the step center is also given. The solid lines are the linear regressions used to interpolate the measured values.} \label{d220-vertical}
\end{figure}

\begin{figure}\centering
\includegraphics[width=7.5cm]{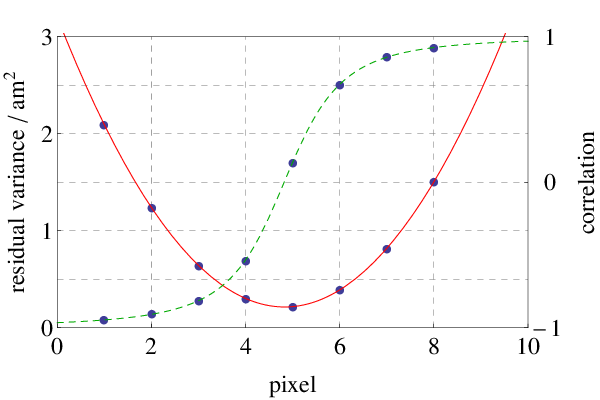}
\caption{Expected (red solid line) and observed (bullets) variances of the residuals from the line best fitting the $\dd$ values interpolated at different detector pixels. Expected (green dashed line) and observed (bullets) correlations between the same residuals and the residuals from the line best fitting the analyser pitch.} \label{correlation}
\end{figure}

\subsection{Abbe's error}\label{Abbe-section}
The Abbe error refers to the difference, $\hat{\mathbf{s}}\cdot(\mathbf{b}\times \bm{\Omega}) = \mathbf{b}\cdot( \bm{\Omega}\times\hat{\mathbf{s}}) = \bm{\Omega}\cdot(\hat{\mathbf{s}}\times\mathbf{b})$, of the displacements sensed by the laser and x-ray interferometers, where $\mathbf{b}$ is the interferometer offset and $\bm{\Omega}$ and $\hat{\mathbf{s}}$ are the rotation and movement-direction of the analyser.

As regards $\bm{\Omega}\times\hat{\mathbf{s}}$, it was zeroed to within 1 nrad (see Sec.\ \ref{alignment-section}) by servoing the motion so as the signals of the angle interferometer -- the phase differences between the vertical and horizontal quadrants of the interference pattern -- are null.

As regards $\hat{\mathbf{s}}\times\mathbf{b}$, the vertical offset was nullified by carrying out off-line measurements of the variations of the x-ray fringe phase in different detector pixels while the pitch component of $\bm{\Omega}$ is purposely changed while keeping the analyser displacement null. As shown in Fig.\ \ref{Abbe}, we identified the virtual pixel having a zero offset to within a 0.1 mm uncertainty. The horizontal offset was set to zero to within the same uncertainty by rotating the analyser about the vertical and by shifting horizontally the laser beam until no phase variation is detected, as shown in Fig.\ \ref{Abbe}.

The angle interferometer and attitude control display imperfections; we took advantage of the resulting pitch noise -- which is shown in Fig.\ \ref{pitch-2} -- to check the vertical offset by data analysis. The eight detector-pixels have a linearly increasing offset and, as shown in Fig.\ \ref{d220-vertical}, the linear regressions of the $\dd$ values intersect, ideally, in the pixel having a null offset. Since the $\dd$ value in this pixel is insensitive to the pitch noise, the best way to find it is to look at the minimum variance of the residuals from the best-fit lines of the values interpolated in each detector pixel or, which is the same, at the zero-correlation between the same residuals and the residuals from the best-fit line of the pitch noise. Examples of the residual variance and correlation are shown in Fig.\ \ref{correlation}. In half of the surveys the null offset is located in a pixel that differs from where the phase variation of the x-ray fringe was found -- in previous off-line experiments -- insensitive to the analyser pitch rotation. After the projection on the analyser, the maximum differences are 0.5 mm. Since the x-ray/optical interferometer rests on a double anti-vibration system (a table supported by air springs, in turn, mounted on a giant pendulum), we explained these shifts by instabilities of the relative levelling between the x-ray source and interferometer which affect the pixel looking at the analyser point having zero vertical offset.

We trusted the zero-offset pixel identified by the data analysis. Since we carried out this analysis after the interferometer was removed from the apparatus and, consequently, we did not investigate experimentally the problem, we assumed a null-pixel error uniform in the $[-0.5,+0.5]$ mm interval with an uncertainty of 0.29 mm. By combining this uncertainty with the uncertainty of the zeroing of the horizontal offset component, 0.1 mm, and parasitic rotation, 2 nrad, we estimated that the Abbe error was nullified to within a total uncertainty of $6.11\times 10^{-10}\dd$.

\begin{figure}\centering
\includegraphics[width=6cm]{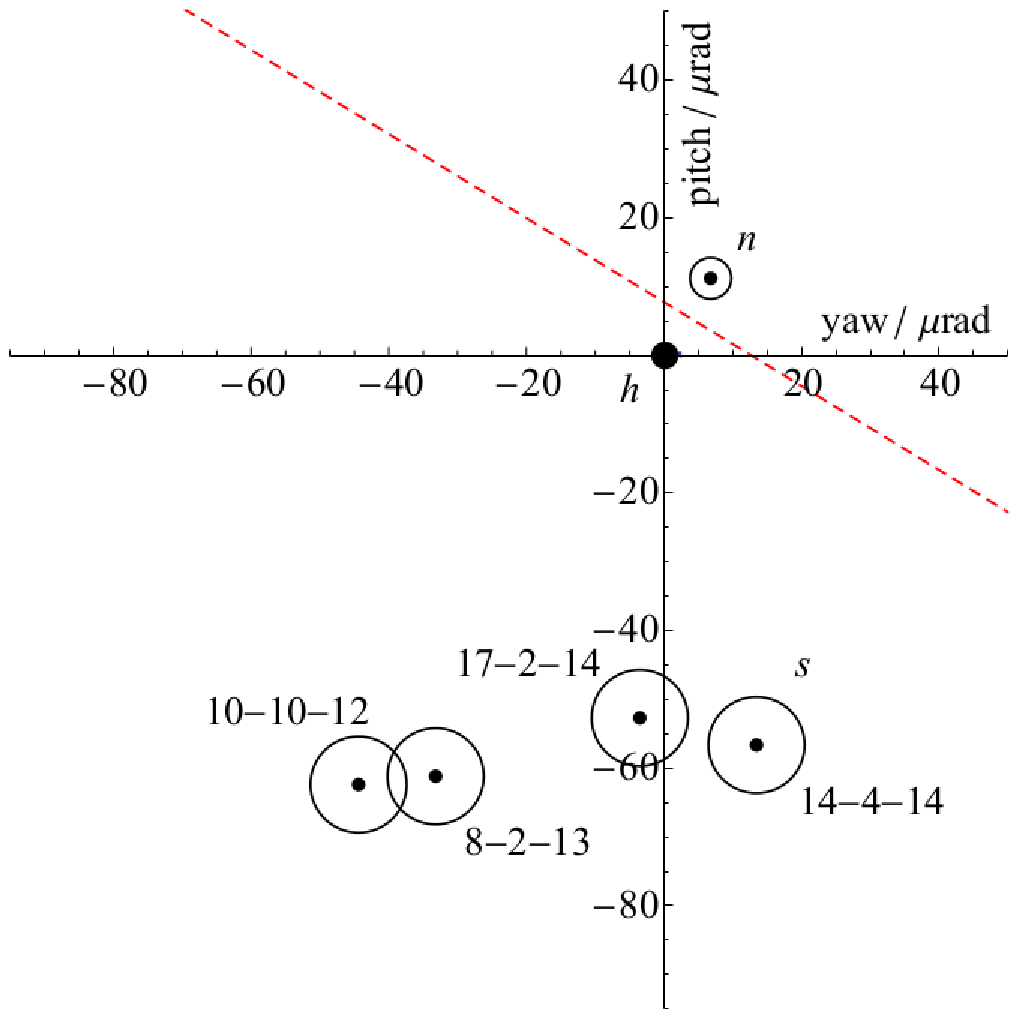}
\caption{Relative directions of the reciprocal vector $\mathbf{h}$ of the diffracting planes, unit normal to the analyser $\mathbf{n}$ (front mirror, with reference to the obverse lay-out of the analyser), and displacement direction $\mathbf{s}$. The dates indicate the $\mathbf{s}$ checks; after 2014/4/14 the analyser was removed and realigned in reverse. The circles indicate the measurement uncertainties. The dashed line is the locus of the $\mathbf{s}$ direction where the projection error is null.} \label{alignment}
\end{figure}

\subsection{Movement direction}
The analyser moves orthogonally to the front mirror of a trihedron; straightness errors are nullified to within nanometers by servoing the motion with the signals of capacitive transducers that sense the transverse displacements of the trihedron top- and side-face.

The misalignment between the optical and x-ray interferometers causes them to measure different components of the displacement. With $\mathbf{s}$ indicating the displacement, the x-ray interferometer senses $\mathbf{s}\cdot\hat{\mathbf{h}}$, where $\hat{\mathbf{h}}$ is the unit normal to the diffracting planes, whereas the optical interferometer senses $\mathbf{s}\cdot\hat{\mathbf{n}}$, where $\hat{\mathbf{n}}$ is the unit normal to the analyser, see Sec.\ \ref{alignment-section}. The difference, $\mathbf{s}\cdot(\hat{\mathbf{n}}-\hat{\mathbf{h}})$, is null when the displacement is orthogonal to $(\hat{\mathbf{n}}-\hat{\mathbf{h}})$, that is, when it bisects the angle formed by $\hat{\mathbf{n}}$ and $\hat{\mathbf{h}}$. The lattice constant is linked to the measured $m\lambda/(2n)$ ratio by
\begin{equation}\label{direction}
 \dd = \frac{m\lambda}{2n} \left[ 1 + \hat{\mathbf{s}}\cdot(\hat{\mathbf{n}}-\hat{\mathbf{h}}) \right] .
\end{equation}

The two (front and rear) analyser-mirrors are polished parallel to the \{220\} planes. The residual misalignments, $|\hat{\mathbf{n}}-\hat{\mathbf{h}}|=13.7(1.2)$ $\mu$rad and $|\hat{\mathbf{n}}-\hat{\mathbf{h}}| = 10.8(1.2)$ $\mu$rad for the front and rear mirrors, with reference to the obverse lay-out of the analyser, were estimated by a least-squares adjustment of the misalignment between x-ray and light reflections on the mirrors and lattice planes,\cite{Mana:1999b} the phase shift of the x-ray fringes when the analyser motion lies in the mirror planes, and the measured angle between the two mirrors. In order to calculate the relevant correction and the associated uncertainty, the angle between the trihedron and analyser was periodically measured to within a $10$ $\mu$rad uncertainty; examples of the measurement results are shown in Fig.\ \ref{alignment}.

\subsection{Analyser temperature}\label{temperature}
The analyser temperature is measured by a capsule standard Pt resistance thermometer inserted into a well in a copper block in thermal contact with the crystal. Resistance measurement were carried out by a FLUKE 1595A Super-Thermometer; to minimize the self-heating, the measurement current was 0.3 mA. Each temperature datum is the average of 15 measurements pairs, carried out with both positive and negative currents and integrated over 30 s. The 1595A linearity was checked by a resistor network made in such a way that the voltages across any number of resistors in a resistor series are read to get four-terminal values interrelated by the formula for the series connection.\cite{Massa:2013} The test showed that linearity is better than 10 $\mu\Omega$ -- corresponding to 25 $\mu$K -- for resistance measurements from 90 $\Omega$ to 120 $\Omega$.

The lattice constant measurements were carried out from February to June 2014; on April 08, the thermometer was calibrated {\it in situ} -- that is, by moving the thermometer from the apparatus to the fixed-point cells without changing the measuring chain and cables. We extrapolated the resistance readings to a zero current and corrected the cell temperatures for the immersion depth and hydrostatic pressure.

The temperature measurements require sub-mK accuracies and any difference between the temperature scales used to extrapolate the molar volume and lattice constant to 20 $^\circ$C must be excluded or identified. Consequently, on April, our fixed-point cells were compared with those of the Physikalish Technische Bundesanstalt. After the corrections for the immersion depth and hydrostatic pressure were taken into account, the differences of the resistance readings were
\[
 \begin{array}{llrl}
 R_{\rm TPW}(\PTB) - R_{\rm TPW}(\INRIM) &=  &5(3) &\mu\Omega \\
 R_{\rm Ga}(\PTB) - R_{\rm Ga}(\INRIM)   &= &-7(4) &\mu\Omega
 \end{array} .
\]
Unfortunately, it was not possible to investigate the non-uniqueness associated with the readings of the two thermometers at 20 $^\circ$C; it was cautiously set to 0.1 mK.\cite{CCT/08-19/rev}

Taking note of these differences, the uncertainties of the triple point of water and melting point of Ga realisations are irrelevant; the repeatability of the cells is 50 $\mu$K. Owing to the huge data averaging, the noise of the resistance measurements is irrelevant; the stability of the reference 100 $\Omega$ resistor over the two months before and after the calibration is 33 $\mu\Omega$, the linearity of the measurement of the 107 $\Omega$ thermometer-resistance is better than 10 $\mu\Omega$, the measurement non-uniqueness is 0.1 mK. All together, the uncertainty of the temperature measurements, estimated by Monte Carlo simulation, is 0.17 mK.

Each $\dd$ measurement was extrapolated to 20 $^\circ$C according to\cite{Bartl:2009}
\begin{equation}
 \dd(T_0) = \dd(T)\left[1 + \alpha_1(T_0-T) + \alpha_2(T_0-T)^2 \right] ,
\end{equation}
where $T_0 = 20\;^\circ{\rm C}$, $\alpha_1 = 2.5530(12)\times 10^{-6}$ K$^{-1}$, and $\alpha_2 = 4.32(37)\times 10^{-9}$ K$^{-2}$. All measurements were carried out in the temperature range from 19.9 $^\circ$C to 20.3 $^\circ$C; therefore, the average extrapolation uncertainty is $0.24\times 10^{-9}\dd$.

The thermometer self-heating was identified by repeating $\dd$ measurements with varying currents; the relevant correction for the 0.3 mA current is $0.33(3)\times 10^{-9}\dd$, the measured value being smaller than the true one.

The calibration history, dating back to December 2007, shows a linear drift of 14(5) $\mu\Omega$/month or 0.035(13) mK/month that was taken into account to extrapolate the calibration to the actual measurement date.

Eventually, the total uncertainty of the lattice constant extrapolation to 20 $^\circ$C is $0.497\times 10^{-9}\dd$.

\begin{figure}\centering
\includegraphics[width=7.5cm]{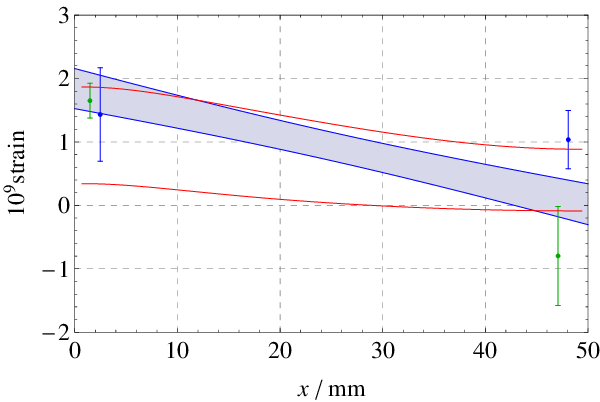}
\caption{Comparison of the finite-element calculation of the minimum (the thermal flux is grounded through the thermometer copper block) and maximum (the thermal flux is grounded through the crystal support points) thermal strains (with respect to the the measured crystal temperature) due to the optical power, 0.75 mW, injected into the analyser by the laser beam (solid lines, red) and the least-squares adjustment (filled area) of the $\dd$ gradients and variations at the crystal ends (dots).} \label{T-strain}
\end{figure}

\subsection{Thermal strain}\label{thermal-strain}
A linear approximation of thermal strain due to the optical power injected into the analyser by the laser beam,
\begin{equation}\label{linear}
 \frac{\Delta\dd}{\dd} = a - b(x-x_0) ,
\end{equation}
where $a=0.948(203)\times 10^{-9}$, $b=0.036(9)\times 10^{-9}$ mm$^{-1}$, and $x_0=22.8$ mm is the center of the survey, was found by a least-squares adjustment of the $\dd$ gradients and the results of repeated $\dd$ measurements carried out with varying optical powers. The comparison of (\ref{linear}) with the numerical calculations of the thermal strain is shown in Fig.\ \ref{T-strain}. To correct for the thermal strain, we trusted the value given by (\ref{linear}), but increased its uncertainty to the one half of the $1.28\times 10^{-9}$ gap between the minimum and maximum strain predicted by the numerical calculation. Therefore, the interpolated value at 22.8 mm was reduced by $0.948(640)\times 10^{-9}\dd$. More details about the numerical and experimental investigations of the analyser response to the thermal load will be given in a separate paper.

\begin{figure}\centering
\includegraphics[width=7.5cm]{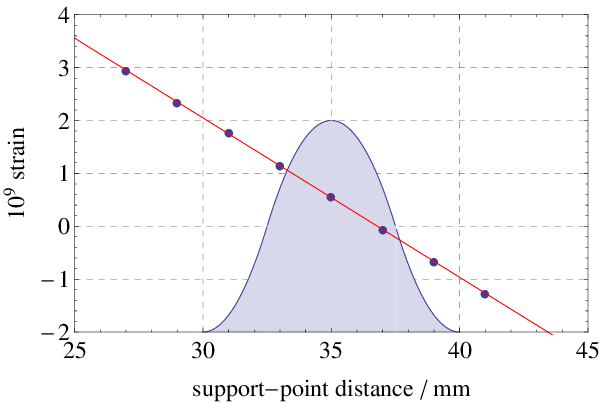}
\caption{Residual mean self-weigh strain of the analyser calculated at 21 mm from the base as a function of the distance between the support points. The filled curve is the distance probability-distribution, given three contact points uniformly distributed in the $(5\times 5)$ mm$^2$ support areas.} \label{strain}
\end{figure}

\subsection{Self-weight deformation}
To average the $\dd$ measurements over the largest crystal part, a long analyser has been used. The simulation of the gravitational bending allowed the analyser to be optimally designed, the residual lattice strain to be predicted, and the contribution of the self-weight deformation to the uncertainty budget estimated.\cite{Mana:2009,Ferroglio:2011} The simulation purposes were to find the maximum height of the analyser lamella consistent with a non-strained lattice and the support points minimizing bending or sagging. In order to estimate the necessary correction and its uncertainty, the residual strain at the 21 mm height was recalculated for points randomly located inside the $(5\times 5)$ mm$^2$ support areas. The results show that the mean strain, which is shown in Fig.\ \ref{strain}, depends only on the distance between the support points. Since the simulation indicates that the everted-inverted transition occurs with a slightly expanded lattice, we reduced the measured values by $0.543(377)\times 10^{-9}\dd$ -- which relates with the distance distribution shown in Fig.\ \ref{strain}.

\subsection{Aberrations of the x-ray interferometer}
Geometric aberrations contribute to the phase of the x-ray fringes by less than $0.01\dd$ per 1 $\mu$m changes of the analyser thickness or focusing.\cite{Mana:1997a,Mana:1997b} The root-mean-square roughness of the analyser surfaces is less than 1 $\mu$m, with its main components in the neighbour of the 0.1 mm wavelength. The effect of the surface roughness -- included the large local variations of the angle with respect to the diffracting planes -- and of local surface strain, if any, were washed out by the survey averaging.

The linear gradient of the mean analyser thickness over the 46 mm measurement distance is less than 10 $\mu$m and contributes by less than $0.4 \times 10^{-9}$ to the $\dd$ measurement. This error is nullified by repeating the measurement after a 180$^\circ$ rotation of the analyser and by averaging the results. If the crystal displacement does not lie in the mean surface of the analyser, the interferometer defocuses. The out-of-plane angle is less than 2 $\mu$m/cm which corresponds to a zero-mean uniform error having 0.23 nm/m standard deviation.

A stress exists in the crystal surfaces even if the bulk material is stress-free. This problem was investigated by Quagliotti {\it et. at}\cite{Quagliotti:2013} by using an elastic-film model to provide a surface load in a finite element analysis. The study showed that, if the tensile stress is 1 N/m, the measured lattice spacing is $6\times 10^{-9}\dd$  smaller than the value in an unstrained crystal. Literature values of the (001) surface-stress obtained from {\it ab initio} and molecular dynamics calculations are given by Quagliotti {\it et. at};\cite{Quagliotti:2013} the stress of the (110) surface is expected to be 60\% smaller. Owing to the value and sign scatters of the literature data, we do not propose a correction and associate with a null stress an uncertainty of 0.1 N/m. Therefore, the relevant contribution to the lattice constant uncertainty is $0.6\times 10^{-9} \dd$. Further experimental investigations and atomistic calculations are under way to confirm that surface stress effects are irrelevant or to quantify and correct for them.\cite{Melis:2014}

\begin{figure}\centering
\includegraphics[width=7.5cm]{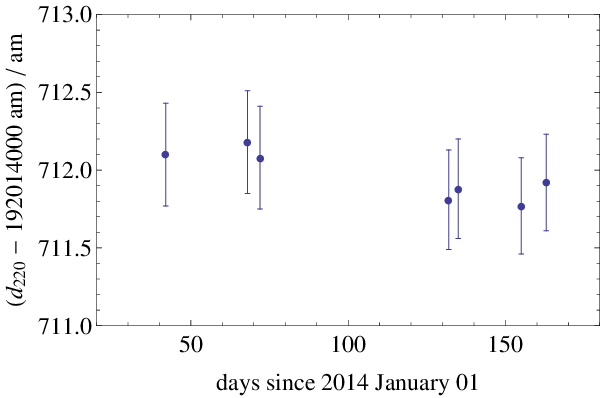}
\caption{Measured $\dd$ values. Each measurement is the mean, after eliminating the outliers and correcting for the thermal strain, of a survey of a 46 mm long crystal part.} \label{d220}
\end{figure}

\section{Measurement results}
Three $\dd$ surveys were made on February 12 and March 10 and 14 with the analyser in the head-on lay-out; four were made on May 14 and 15 and June 05 and 12 with the analyser in the back orientation. Examples are given in Figs.\ \ref{d220a} and \ref{d220b}. Next, after eliminating the outliers and correcting for the thermal strain, each $\dd$ profile was averaged to obtain the mean lattice spacing.

\begin{figure}\centering
\includegraphics[width=7.5cm]{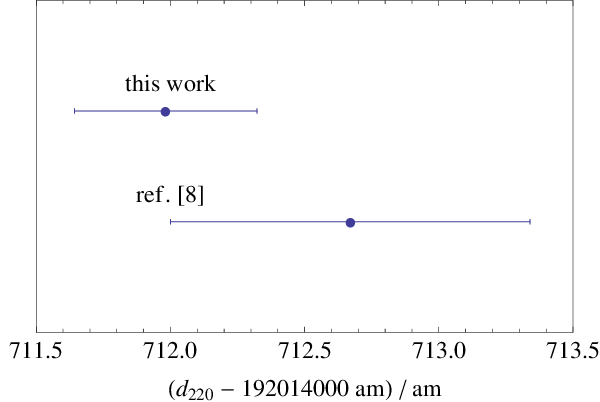}
\caption{Comparison of the $\dd$ values (\ref{value}) and (\ref{value:2011}).} \label{d220-2014}
\end{figure}

The results are shown in Fig.\ \ref{d220}; an example of the error budget is given in Table \ref{table:1}. With respect to our previous measurement, in addition to the reduction of the total uncertainty, the Table \ref{table:1} shows a significant redistribution of the uncertainty contributions. The values belonging to each obverse/reverse faces set are significantly correlated; this explains the repeatability, which is much better than the uncertainty.

The analyser reversal required a full realignment of the two interferometers. Therefore, only the wavelength and temperature uncertainties, laser-beam diffraction, self-weigh deformation, and aberrations of the x-ray interferometer combine in the same way; the remaining contributions to the total uncertainty are largely independent.

Figure \ref{d220} shows a difference between the values measured with the analyser mounted in the head-on and inverted lay-outs. We are not yet able to say if this difference is real, e.g., due to a different physical and/or chemical structure of the analyser surfaces, or if it indicates that the control of the systematic errors is less good than what we estimated. The first option is supported by the repeatable observation of different obverse- and reverse-profile, which was statistically anticipated by Massa {\it et al.}.\cite{Massa:2011a} The second is supported by the fact that a difference between the obverse and reverse mean-values of $\dd$ was not reported.

The final measured value,
\begin{equation}\label{value}
 \dd(2014)=192014711.98(34)\; {\rm am},
\end{equation}
at 20 $^\circ$C and 0 Pa, is the mean of the data in Fig.\ \ref{d220}. The relative uncertainty is 1.75 nm/m. In the average, we did not take the data uncertainty and correlation into account, but, to avoid that the different number of obverse/reverse surveys biases the result, firstly, we averaged the obverse and reverse data and, subsequently, averaged the two results. To be conservative, we associated to the mean the worst uncertainty of the input data.

\begin{table}
\caption{\label{table:1} Relative correction and uncertainty, in parts per $10^9$, of the 2014/02/12 $\dd$ value.}
\begin{ruledtabular}
\begin{tabular}{@{}lrr}
Contribution           &Correction &Uncertainty \\
data averaging         & 0.000       & 0.722\\
wavelength             & $-0.058$    & 0.033\\
laser beam diffraction & 3.978       & 0.597\\
laser beam alignment   & $-0.110$    & 0.480\\
beam walks             & 0.000       & 0.577\\
Abbe's errors          & 0.000       & 0.611\\
movement direction     & 0.699       & 0.214\\
temperature            & $-0.500$    & 0.497\\
thermal strain         & $-0.948$    & 0.641\\
self-weigh             & $-0.543$    & 0.377\\
aberrations            & 0.000       & 0.642\\
total                  & 2.52        & 1.75\\
\end{tabular}
\end{ruledtabular}
\end{table}

As shown in Fig.\ \ref{d220-2014}, (\ref{value}) is slightly smaller than
\begin{equation}\label{value:2011}
 \dd(2011)=192014712.67(67)\; {\rm am}
\end{equation}
given by Massa {\it et al.}.\cite{Massa:2011a} A reasons might be a positive bias of the correction for diffraction applied to (\ref{value:2011}), as shown in Fig.\ \ref{diffraction}. In addition, in (\ref{value:2011}), the pointing stability of the laser beam was not monitored on-line and, in retrospect, we might have overestimated the pointing error. Since a pointing error causes a $\chi^2$ measurement-underestimate, we consistently -- but, perhaps incorrectly -- applied a relatively large positive correction. Eventually, the reanalysis of the effect of a non-orthogonal incidence of the laser beam on the analyser mirror in Sec.\ \ref{alignment-section} shows that wedged optics in the beam path combine with a wrong pointing to originate a positive error. Therefore, contrary to what we did in the past, the correction for the laser beam alignment in Table \ref{table:1} is negative. {\colr To estimate the correlation between the (\ref{value}) and (\ref{value:2011}) values, a detailed analysis of the similarities and differences of our present and past measurements  is under way; the results will be given in a separate paper.}

\section{Conclusions}
We re-checked the measurement uncertainty of the lattice parameter of the $\eSi$ crystal used to determine the Avogadro constant. Additional measurements and stress-tests were carried out by using of an upgraded measurement apparatus. No error was identified; this work confirms the value given by Massa {\it et al.}\cite{Massa:2011a} and yields an additional result having a reduced uncertainty.

\begin{acknowledgments}
This work was jointly funded by the European Me\-trology Research Pro\-gramme (EMRP) par\-ti\-ci\-pa\-ting coun\-tries within the European Association of National Metrology Institutes (EURAMET), the European Union, and the Italian ministry of education, university, and research (awarded project P6-2013, implementation of the new SI).
\end{acknowledgments}


\begin{thebibliography}{36}%
\makeatletter
\providecommand \@ifxundefined [1]{%
 \@ifx{#1\undefined}
}%
\providecommand \@ifnum [1]{%
 \ifnum #1\expandafter \@firstoftwo
 \else \expandafter \@secondoftwo
 \fi
}%
\providecommand \@ifx [1]{%
 \ifx #1\expandafter \@firstoftwo
 \else \expandafter \@secondoftwo
 \fi
}%
\providecommand \natexlab [1]{#1}%
\providecommand \enquote  [1]{``#1''}%
\providecommand \bibnamefont  [1]{#1}%
\providecommand \bibfnamefont [1]{#1}%
\providecommand \citenamefont [1]{#1}%
\providecommand \href@noop [0]{\@secondoftwo}%
\providecommand \href [0]{\begingroup \@sanitize@url \@href}%
\providecommand \@href[1]{\@@startlink{#1}\@@href}%
\providecommand \@@href[1]{\endgroup#1\@@endlink}%
\providecommand \@sanitize@url [0]{\catcode `\\12\catcode `\$12\catcode
  `\&12\catcode `\#12\catcode `\^12\catcode `\_12\catcode `\%12\relax}%
\providecommand \@@startlink[1]{}%
\providecommand \@@endlink[0]{}%
\providecommand \url  [0]{\begingroup\@sanitize@url \@url }%
\providecommand \@url [1]{\endgroup\@href {#1}{\urlprefix }}%
\providecommand \urlprefix  [0]{URL }%
\providecommand \Eprint [0]{\href }%
\providecommand \doibase [0]{http://dx.doi.org/}%
\providecommand \selectlanguage [0]{\@gobble}%
\providecommand \bibinfo  [0]{\@secondoftwo}%
\providecommand \bibfield  [0]{\@secondoftwo}%
\providecommand \translation [1]{[#1]}%
\providecommand \BibitemOpen [0]{}%
\providecommand \bibitemStop [0]{}%
\providecommand \bibitemNoStop [0]{.\EOS\space}%
\providecommand \EOS [0]{\spacefactor3000\relax}%
\providecommand \BibitemShut  [1]{\csname bibitem#1\endcsname}%
\let\auto@bib@innerbib\@empty
\bibitem [{\citenamefont {Bettin}\ \emph {et~al.}(2013)\citenamefont {Bettin},
  \citenamefont {Fujii}, \citenamefont {Man}, \citenamefont {Mana},
  \citenamefont {Massa},\ and\ \citenamefont {Picard}}]{Bettin:2013}%
  \BibitemOpen
  \bibfield  {author} {\bibinfo {author} {\bibfnamefont {H.}~\bibnamefont
  {Bettin}}, \bibinfo {author} {\bibfnamefont {K.}~\bibnamefont {Fujii}},
  \bibinfo {author} {\bibfnamefont {J.}~\bibnamefont {Man}}, \bibinfo {author}
  {\bibfnamefont {G.}~\bibnamefont {Mana}}, \bibinfo {author} {\bibfnamefont
  {E.}~\bibnamefont {Massa}}, \ and\ \bibinfo {author} {\bibfnamefont
  {A.}~\bibnamefont {Picard}},\ }\href@noop {} {\bibfield  {journal} {\bibinfo
  {journal} {Ann.\ Phys}\ }\textbf {\bibinfo {volume} {525}},\ \bibinfo {pages}
  {680} (\bibinfo {year} {2013})}\BibitemShut {NoStop}%
\bibitem [{\citenamefont {Mana}\ and\ \citenamefont {Massa}(2012)}]{Mana:2012}%
  \BibitemOpen
  \bibfield  {author} {\bibinfo {author} {\bibfnamefont {G.}~\bibnamefont
  {Mana}}\ and\ \bibinfo {author} {\bibfnamefont {E.}~\bibnamefont {Massa}},\
  }\href@noop {} {\bibfield  {journal} {\bibinfo  {journal} {Rivista del Nuovo
  Cimento}\ }\textbf {\bibinfo {volume} {35}},\ \bibinfo {pages} {353}
  (\bibinfo {year} {2012})}\BibitemShut {NoStop}%
\bibitem [{\citenamefont {Andreas}\ \emph
  {et~al.}(2011{\natexlab{a}})\citenamefont {Andreas}, \citenamefont {Azuma},
  \citenamefont {Bartl}, \citenamefont {Becker}, \citenamefont {Bettin},
  \citenamefont {Borys}, \citenamefont {Busch}, \citenamefont {Gray},
  \citenamefont {Fuchs}, \citenamefont {Fujii}, \citenamefont {Fujimoto},
  \citenamefont {Kessler}, \citenamefont {Krumrey}, \citenamefont {Kuetgens},
  \citenamefont {Kuramoto}, \citenamefont {Mana}, \citenamefont {Manson},
  \citenamefont {Massa}, \citenamefont {Mizushima}, \citenamefont {Nicolaus},
  \citenamefont {Picard}, \citenamefont {Pramann}, \citenamefont {Rienitz},
  \citenamefont {Schiel}, \citenamefont {Valkiers},\ and\ \citenamefont
  {Waseda}}]{Andreas:2011a}%
  \BibitemOpen
  \bibfield  {author} {\bibinfo {author} {\bibfnamefont {B.}~\bibnamefont
  {Andreas}}, \bibinfo {author} {\bibfnamefont {Y.}~\bibnamefont {Azuma}},
  \bibinfo {author} {\bibfnamefont {G.}~\bibnamefont {Bartl}}, \bibinfo
  {author} {\bibfnamefont {P.}~\bibnamefont {Becker}}, \bibinfo {author}
  {\bibfnamefont {H.}~\bibnamefont {Bettin}}, \bibinfo {author} {\bibfnamefont
  {M.}~\bibnamefont {Borys}}, \bibinfo {author} {\bibfnamefont
  {I.}~\bibnamefont {Busch}}, \bibinfo {author} {\bibfnamefont
  {M.}~\bibnamefont {Gray}}, \bibinfo {author} {\bibfnamefont {P.}~\bibnamefont
  {Fuchs}}, \bibinfo {author} {\bibfnamefont {K.}~\bibnamefont {Fujii}},
  \bibinfo {author} {\bibfnamefont {H.}~\bibnamefont {Fujimoto}}, \bibinfo
  {author} {\bibfnamefont {E.}~\bibnamefont {Kessler}}, \bibinfo {author}
  {\bibfnamefont {M.}~\bibnamefont {Krumrey}}, \bibinfo {author} {\bibfnamefont
  {U.}~\bibnamefont {Kuetgens}}, \bibinfo {author} {\bibfnamefont
  {N.}~\bibnamefont {Kuramoto}}, \bibinfo {author} {\bibfnamefont
  {G.}~\bibnamefont {Mana}}, \bibinfo {author} {\bibfnamefont {P.}~\bibnamefont
  {Manson}}, \bibinfo {author} {\bibfnamefont {E.}~\bibnamefont {Massa}},
  \bibinfo {author} {\bibfnamefont {S.}~\bibnamefont {Mizushima}}, \bibinfo
  {author} {\bibfnamefont {A.}~\bibnamefont {Nicolaus}}, \bibinfo {author}
  {\bibfnamefont {A.}~\bibnamefont {Picard}}, \bibinfo {author} {\bibfnamefont
  {A.}~\bibnamefont {Pramann}}, \bibinfo {author} {\bibfnamefont
  {O.}~\bibnamefont {Rienitz}}, \bibinfo {author} {\bibfnamefont
  {D.}~\bibnamefont {Schiel}}, \bibinfo {author} {\bibfnamefont
  {S.}~\bibnamefont {Valkiers}}, \ and\ \bibinfo {author} {\bibfnamefont
  {A.}~\bibnamefont {Waseda}},\ }\href@noop {} {\bibfield  {journal} {\bibinfo
  {journal} {Phys.\ Rev.\ Lett.}\ }\textbf {\bibinfo {volume} {106}} (\bibinfo
  {year} {2011}{\natexlab{a}})}\BibitemShut {NoStop}%
\bibitem [{\citenamefont {Andreas}\ \emph
  {et~al.}(2011{\natexlab{b}})\citenamefont {Andreas}, \citenamefont {Azuma},
  \citenamefont {Bartl}, \citenamefont {Becker}, \citenamefont {Bettin},
  \citenamefont {Borys}, \citenamefont {Busch}, \citenamefont {Fuchs},
  \citenamefont {Fujii}, \citenamefont {Fujimoto}, \citenamefont {Kessler},
  \citenamefont {Krumrey}, \citenamefont {Kuetgens}, \citenamefont {Kuramoto},
  \citenamefont {Mana}, \citenamefont {Massa}, \citenamefont {Mizushima},
  \citenamefont {Nicolaus}, \citenamefont {Picard}, \citenamefont {Pramann},
  \citenamefont {Rienitz}, \citenamefont {Schiel}, \citenamefont {Valkiers},
  \citenamefont {Waseda},\ and\ \citenamefont {Zakel}}]{Andreas:2011b}%
  \BibitemOpen
  \bibfield  {author} {\bibinfo {author} {\bibfnamefont {B.}~\bibnamefont
  {Andreas}}, \bibinfo {author} {\bibfnamefont {Y.}~\bibnamefont {Azuma}},
  \bibinfo {author} {\bibfnamefont {G.}~\bibnamefont {Bartl}}, \bibinfo
  {author} {\bibfnamefont {P.}~\bibnamefont {Becker}}, \bibinfo {author}
  {\bibfnamefont {H.}~\bibnamefont {Bettin}}, \bibinfo {author} {\bibfnamefont
  {M.}~\bibnamefont {Borys}}, \bibinfo {author} {\bibfnamefont
  {I.}~\bibnamefont {Busch}}, \bibinfo {author} {\bibfnamefont
  {P.}~\bibnamefont {Fuchs}}, \bibinfo {author} {\bibfnamefont
  {K.}~\bibnamefont {Fujii}}, \bibinfo {author} {\bibfnamefont
  {H.}~\bibnamefont {Fujimoto}}, \bibinfo {author} {\bibfnamefont
  {E.}~\bibnamefont {Kessler}}, \bibinfo {author} {\bibfnamefont
  {M.}~\bibnamefont {Krumrey}}, \bibinfo {author} {\bibfnamefont
  {U.}~\bibnamefont {Kuetgens}}, \bibinfo {author} {\bibfnamefont
  {N.}~\bibnamefont {Kuramoto}}, \bibinfo {author} {\bibfnamefont
  {G.}~\bibnamefont {Mana}}, \bibinfo {author} {\bibfnamefont {E.}~\bibnamefont
  {Massa}}, \bibinfo {author} {\bibfnamefont {S.}~\bibnamefont {Mizushima}},
  \bibinfo {author} {\bibfnamefont {A.}~\bibnamefont {Nicolaus}}, \bibinfo
  {author} {\bibfnamefont {A.}~\bibnamefont {Picard}}, \bibinfo {author}
  {\bibfnamefont {A.}~\bibnamefont {Pramann}}, \bibinfo {author} {\bibfnamefont
  {O.}~\bibnamefont {Rienitz}}, \bibinfo {author} {\bibfnamefont
  {D.}~\bibnamefont {Schiel}}, \bibinfo {author} {\bibfnamefont
  {S.}~\bibnamefont {Valkiers}}, \bibinfo {author} {\bibfnamefont
  {A.}~\bibnamefont {Waseda}}, \ and\ \bibinfo {author} {\bibfnamefont
  {S.}~\bibnamefont {Zakel}},\ }\href@noop {} {\bibfield  {journal} {\bibinfo
  {journal} {Metrologia}\ }\textbf {\bibinfo {volume} {48}},\ \bibinfo {pages}
  {S1} (\bibinfo {year} {2011}{\natexlab{b}})}\BibitemShut {NoStop}%
\bibitem [{\citenamefont {Azuma}\ \emph {et~al.}(2015)\citenamefont {Azuma},
  \citenamefont {Barat}, \citenamefont {Bartl}, \citenamefont {Bettin},
  \citenamefont {Borys}, \citenamefont {Busch}, \citenamefont {Cibik},
  \citenamefont {D’Agostino}, \citenamefont {Fujii}, \citenamefont
  {Fujimoto}, \citenamefont {Hioki}, \citenamefont {Krumrey}, \citenamefont
  {Kuetgens}, \citenamefont {Kuramoto}, \citenamefont {Mana}, \citenamefont
  {Massa}, \citenamefont {Mee\ss}, \citenamefont {Mizushima}, \citenamefont
  {Narukawa}, \citenamefont {Nicolaus}, \citenamefont {Pramann}, \citenamefont
  {Rabb}, \citenamefont {Rienitz}, \citenamefont {Sasso}, \citenamefont
  {Stock}, \citenamefont {Vocke}, \citenamefont {Waseda}, \citenamefont
  {Wundrack},\ and\ \citenamefont {Zakel}}]{Azuma:2015}%
  \BibitemOpen
  \bibfield  {author} {\bibinfo {author} {\bibfnamefont {Y.}~\bibnamefont
  {Azuma}}, \bibinfo {author} {\bibfnamefont {P.}~\bibnamefont {Barat}},
  \bibinfo {author} {\bibfnamefont {G.}~\bibnamefont {Bartl}}, \bibinfo
  {author} {\bibfnamefont {H.}~\bibnamefont {Bettin}}, \bibinfo {author}
  {\bibfnamefont {M.}~\bibnamefont {Borys}}, \bibinfo {author} {\bibfnamefont
  {I.}~\bibnamefont {Busch}}, \bibinfo {author} {\bibfnamefont
  {L.}~\bibnamefont {Cibik}}, \bibinfo {author} {\bibfnamefont
  {G.}~\bibnamefont {D’Agostino}}, \bibinfo {author} {\bibfnamefont
  {K.}~\bibnamefont {Fujii}}, \bibinfo {author} {\bibfnamefont
  {H.}~\bibnamefont {Fujimoto}}, \bibinfo {author} {\bibfnamefont
  {A.}~\bibnamefont {Hioki}}, \bibinfo {author} {\bibfnamefont
  {M.}~\bibnamefont {Krumrey}}, \bibinfo {author} {\bibfnamefont
  {U.}~\bibnamefont {Kuetgens}}, \bibinfo {author} {\bibfnamefont
  {N.}~\bibnamefont {Kuramoto}}, \bibinfo {author} {\bibfnamefont
  {G.}~\bibnamefont {Mana}}, \bibinfo {author} {\bibfnamefont {E.}~\bibnamefont
  {Massa}}, \bibinfo {author} {\bibfnamefont {R.}~\bibnamefont {Mee\ss}},
  \bibinfo {author} {\bibfnamefont {S.}~\bibnamefont {Mizushima}}, \bibinfo
  {author} {\bibfnamefont {T.}~\bibnamefont {Narukawa}}, \bibinfo {author}
  {\bibfnamefont {A.}~\bibnamefont {Nicolaus}}, \bibinfo {author}
  {\bibfnamefont {A.}~\bibnamefont {Pramann}}, \bibinfo {author} {\bibfnamefont
  {S.~A.}\ \bibnamefont {Rabb}}, \bibinfo {author} {\bibfnamefont
  {O.}~\bibnamefont {Rienitz}}, \bibinfo {author} {\bibfnamefont
  {C.}~\bibnamefont {Sasso}}, \bibinfo {author} {\bibfnamefont
  {M.}~\bibnamefont {Stock}}, \bibinfo {author} {\bibfnamefont {R.~D.~J.}\
  \bibnamefont {Vocke}}, \bibinfo {author} {\bibfnamefont {A.}~\bibnamefont
  {Waseda}}, \bibinfo {author} {\bibfnamefont {S.}~\bibnamefont {Wundrack}}, \
  and\ \bibinfo {author} {\bibfnamefont {S.}~\bibnamefont {Zakel}},\
  }\href@noop {} {\bibfield  {journal} {\bibinfo  {journal} {Metrologia}\
  }\textbf {\bibinfo {volume} {52}},\ \bibinfo {pages} {accepeted} (\bibinfo
  {year} {2015})}\BibitemShut {NoStop}%
\bibitem [{\citenamefont {Steiner}\ \emph {et~al.}(2007)\citenamefont
  {Steiner}, \citenamefont {Williams}, \citenamefont {Liu},\ and\ \citenamefont
  {Newell}}]{Steiner:2007}%
  \BibitemOpen
  \bibfield  {author} {\bibinfo {author} {\bibfnamefont {R.}~\bibnamefont
  {Steiner}}, \bibinfo {author} {\bibfnamefont {E.}~\bibnamefont {Williams}},
  \bibinfo {author} {\bibfnamefont {R.}~\bibnamefont {Liu}}, \ and\ \bibinfo
  {author} {\bibfnamefont {D.}~\bibnamefont {Newell}},\ }\href@noop {}
  {\bibfield  {journal} {\bibinfo  {journal} {IEEE Trans.\ Instrum.\ Meas.}\
  }\textbf {\bibinfo {volume} {56}},\ \bibinfo {pages} {592} (\bibinfo {year}
  {2007})}\BibitemShut {NoStop}%
\bibitem [{\citenamefont {Sanchez}\ \emph {et~al.}(2014)\citenamefont
  {Sanchez}, \citenamefont {Wood}, \citenamefont {Green}, \citenamefont
  {Liard},\ and\ \citenamefont {Inglis}}]{Sanchez:2014}%
  \BibitemOpen
  \bibfield  {author} {\bibinfo {author} {\bibfnamefont {C.~A.}\ \bibnamefont
  {Sanchez}}, \bibinfo {author} {\bibfnamefont {B.~M.}\ \bibnamefont {Wood}},
  \bibinfo {author} {\bibfnamefont {R.~G.}\ \bibnamefont {Green}}, \bibinfo
  {author} {\bibfnamefont {J.~O.}\ \bibnamefont {Liard}}, \ and\ \bibinfo
  {author} {\bibfnamefont {D.}~\bibnamefont {Inglis}},\ }\href@noop {}
  {\bibfield  {journal} {\bibinfo  {journal} {Metrologia}\ }\textbf {\bibinfo
  {volume} {51}},\ \bibinfo {pages} {S5} (\bibinfo {year} {2014})}\BibitemShut
  {NoStop}%
\bibitem [{\citenamefont {Schlamminger}\ \emph {et~al.}(2014)\citenamefont
  {Schlamminger}, \citenamefont {Haddad}, \citenamefont {Seifert},
  \citenamefont {Chao}, \citenamefont {Newell}, \citenamefont {Liu},
  \citenamefont {Steiner},\ and\ \citenamefont {Pratt}}]{Schlamminger:2014}%
  \BibitemOpen
  \bibfield  {author} {\bibinfo {author} {\bibfnamefont {S.}~\bibnamefont
  {Schlamminger}}, \bibinfo {author} {\bibfnamefont {D.}~\bibnamefont
  {Haddad}}, \bibinfo {author} {\bibfnamefont {F.}~\bibnamefont {Seifert}},
  \bibinfo {author} {\bibfnamefont {L.~S.}\ \bibnamefont {Chao}}, \bibinfo
  {author} {\bibfnamefont {D.~B.}\ \bibnamefont {Newell}}, \bibinfo {author}
  {\bibfnamefont {R.}~\bibnamefont {Liu}}, \bibinfo {author} {\bibfnamefont
  {R.~L.}\ \bibnamefont {Steiner}}, \ and\ \bibinfo {author} {\bibfnamefont
  {J.~R.}\ \bibnamefont {Pratt}},\ }\href@noop {} {\bibfield  {journal}
  {\bibinfo  {journal} {Metrologia}\ }\textbf {\bibinfo {volume} {51}},\
  \bibinfo {pages} {S15} (\bibinfo {year} {2014})}\BibitemShut {NoStop}%
\bibitem [{\citenamefont {Massa}\ \emph
  {et~al.}(2011{\natexlab{a}})\citenamefont {Massa}, \citenamefont {Mana},
  \citenamefont {Kuetgens},\ and\ \citenamefont {Ferroglio}}]{Massa:2011a}%
  \BibitemOpen
  \bibfield  {author} {\bibinfo {author} {\bibfnamefont {E.}~\bibnamefont
  {Massa}}, \bibinfo {author} {\bibfnamefont {G.}~\bibnamefont {Mana}},
  \bibinfo {author} {\bibfnamefont {U.}~\bibnamefont {Kuetgens}}, \ and\
  \bibinfo {author} {\bibfnamefont {L.}~\bibnamefont {Ferroglio}},\ }\href@noop
  {} {\bibfield  {journal} {\bibinfo  {journal} {Metrologia}\ }\textbf
  {\bibinfo {volume} {48}},\ \bibinfo {pages} {S37} (\bibinfo {year}
  {2011}{\natexlab{a}})}\BibitemShut {NoStop}%
\bibitem [{\citenamefont {Pramann}\ \emph {et~al.}(2011)\citenamefont
  {Pramann}, \citenamefont {Rienitz}, \citenamefont {Schiel}, \citenamefont
  {Schlote}, \citenamefont {Güttler},\ and\ \citenamefont
  {Valkiers}}]{Pramann:2011}%
  \BibitemOpen
  \bibfield  {author} {\bibinfo {author} {\bibfnamefont {A.}~\bibnamefont
  {Pramann}}, \bibinfo {author} {\bibfnamefont {O.}~\bibnamefont {Rienitz}},
  \bibinfo {author} {\bibfnamefont {D.}~\bibnamefont {Schiel}}, \bibinfo
  {author} {\bibfnamefont {J.}~\bibnamefont {Schlote}}, \bibinfo {author}
  {\bibfnamefont {B.}~\bibnamefont {Güttler}}, \ and\ \bibinfo {author}
  {\bibfnamefont {S.}~\bibnamefont {Valkiers}},\ }\href@noop {} {\bibfield
  {journal} {\bibinfo  {journal} {Metrologia}\ }\textbf {\bibinfo {volume}
  {48}},\ \bibinfo {pages} {S20} (\bibinfo {year} {2011})}\BibitemShut
  {NoStop}%
\bibitem [{\citenamefont {Narukawa}\ \emph {et~al.}(2014)\citenamefont
  {Narukawa}, \citenamefont {Hioki}, \citenamefont {Kuramoto},\ and\
  \citenamefont {Fujii}}]{Narukawa:2014}%
  \BibitemOpen
  \bibfield  {author} {\bibinfo {author} {\bibfnamefont {T.}~\bibnamefont
  {Narukawa}}, \bibinfo {author} {\bibfnamefont {A.}~\bibnamefont {Hioki}},
  \bibinfo {author} {\bibfnamefont {N.}~\bibnamefont {Kuramoto}}, \ and\
  \bibinfo {author} {\bibfnamefont {K.}~\bibnamefont {Fujii}},\ }\href@noop {}
  {\bibfield  {journal} {\bibinfo  {journal} {Metrologia}\ }\textbf {\bibinfo
  {volume} {51}},\ \bibinfo {pages} {161} (\bibinfo {year} {2014})}\BibitemShut
  {NoStop}%
\bibitem [{\citenamefont {Jr}, \citenamefont {Rabb},\ and\ \citenamefont
  {Turk}(2014)}]{Vocke:2014}%
  \BibitemOpen
  \bibfield  {author} {\bibinfo {author} {\bibfnamefont {R.~D.~V.}\
  \bibnamefont {Jr}}, \bibinfo {author} {\bibfnamefont {S.~A.}\ \bibnamefont
  {Rabb}}, \ and\ \bibinfo {author} {\bibfnamefont {G.~C.}\ \bibnamefont
  {Turk}},\ }\href@noop {} {\bibfield  {journal} {\bibinfo  {journal}
  {Metrologia}\ }\textbf {\bibinfo {volume} {51}},\ \bibinfo {pages} {361}
  (\bibinfo {year} {2014})}\BibitemShut {NoStop}%
\bibitem [{\citenamefont {Picard}\ \emph {et~al.}(2011)\citenamefont {Picard},
  \citenamefont {Barat}, \citenamefont {Borys}, \citenamefont {Firlus},\ and\
  \citenamefont {Mizushima}}]{Picard:2011}%
  \BibitemOpen
  \bibfield  {author} {\bibinfo {author} {\bibfnamefont {A.}~\bibnamefont
  {Picard}}, \bibinfo {author} {\bibfnamefont {P.}~\bibnamefont {Barat}},
  \bibinfo {author} {\bibfnamefont {M.}~\bibnamefont {Borys}}, \bibinfo
  {author} {\bibfnamefont {M.}~\bibnamefont {Firlus}}, \ and\ \bibinfo {author}
  {\bibfnamefont {S.}~\bibnamefont {Mizushima}},\ }\href@noop {} {\bibfield
  {journal} {\bibinfo  {journal} {Metrologia}\ }\textbf {\bibinfo {volume}
  {48}},\ \bibinfo {pages} {S112} (\bibinfo {year} {2011})}\BibitemShut
  {NoStop}%
\bibitem [{\citenamefont {Kuramoto}, \citenamefont {Fujii},\ and\ \citenamefont
  {Yamazawa}(2011)}]{Kuramoto:2011}%
  \BibitemOpen
  \bibfield  {author} {\bibinfo {author} {\bibfnamefont {N.}~\bibnamefont
  {Kuramoto}}, \bibinfo {author} {\bibfnamefont {K.}~\bibnamefont {Fujii}}, \
  and\ \bibinfo {author} {\bibfnamefont {K.}~\bibnamefont {Yamazawa}},\
  }\href@noop {} {\bibfield  {journal} {\bibinfo  {journal} {Metrologia}\
  }\textbf {\bibinfo {volume} {48}},\ \bibinfo {pages} {S83} (\bibinfo {year}
  {2011})}\BibitemShut {NoStop}%
\bibitem [{\citenamefont {Bartl}\ \emph {et~al.}(2011)\citenamefont {Bartl},
  \citenamefont {Bettin}, \citenamefont {Krystek}, \citenamefont {Mai},
  \citenamefont {Nicolaus},\ and\ \citenamefont {Peter}}]{Bartl:2011}%
  \BibitemOpen
  \bibfield  {author} {\bibinfo {author} {\bibfnamefont {G.}~\bibnamefont
  {Bartl}}, \bibinfo {author} {\bibfnamefont {H.}~\bibnamefont {Bettin}},
  \bibinfo {author} {\bibfnamefont {M.}~\bibnamefont {Krystek}}, \bibinfo
  {author} {\bibfnamefont {T.}~\bibnamefont {Mai}}, \bibinfo {author}
  {\bibfnamefont {A.}~\bibnamefont {Nicolaus}}, \ and\ \bibinfo {author}
  {\bibfnamefont {A.}~\bibnamefont {Peter}},\ }\href@noop {} {\bibfield
  {journal} {\bibinfo  {journal} {Metrologia}\ }\textbf {\bibinfo {volume}
  {48}},\ \bibinfo {pages} {S96} (\bibinfo {year} {2011})}\BibitemShut
  {NoStop}%
\bibitem [{\citenamefont {Fujimoto}, \citenamefont {Waseda},\ and\
  \citenamefont {Zhang}(2011)}]{Fujimoto:2011}%
  \BibitemOpen
  \bibfield  {author} {\bibinfo {author} {\bibfnamefont {H.}~\bibnamefont
  {Fujimoto}}, \bibinfo {author} {\bibfnamefont {A.}~\bibnamefont {Waseda}}, \
  and\ \bibinfo {author} {\bibfnamefont {X.~W.}\ \bibnamefont {Zhang}},\
  }\href@noop {} {\bibfield  {journal} {\bibinfo  {journal} {Metrologia}\
  }\textbf {\bibinfo {volume} {48}},\ \bibinfo {pages} {S55} (\bibinfo {year}
  {2011})}\BibitemShut {NoStop}%
\bibitem [{\citenamefont {Massa}\ \emph
  {et~al.}(2011{\natexlab{b}})\citenamefont {Massa}, \citenamefont {Mana},
  \citenamefont {Ferroglio}, \citenamefont {Kessler}, \citenamefont {Schiel},\
  and\ \citenamefont {Zakel}}]{Massa:2011b}%
  \BibitemOpen
  \bibfield  {author} {\bibinfo {author} {\bibfnamefont {E.}~\bibnamefont
  {Massa}}, \bibinfo {author} {\bibfnamefont {G.}~\bibnamefont {Mana}},
  \bibinfo {author} {\bibfnamefont {L.}~\bibnamefont {Ferroglio}}, \bibinfo
  {author} {\bibfnamefont {E.~G.}\ \bibnamefont {Kessler}}, \bibinfo {author}
  {\bibfnamefont {D.}~\bibnamefont {Schiel}}, \ and\ \bibinfo {author}
  {\bibfnamefont {S.}~\bibnamefont {Zakel}},\ }\href@noop {} {\bibfield
  {journal} {\bibinfo  {journal} {Metrologia}\ }\textbf {\bibinfo {volume}
  {48}},\ \bibinfo {pages} {S44} (\bibinfo {year}
  {2011}{\natexlab{b}})}\BibitemShut {NoStop}%
\bibitem [{\citenamefont {Zakel}\ \emph {et~al.}(2011)\citenamefont {Zakel},
  \citenamefont {Wundrack}, \citenamefont {Niemann}, \citenamefont {Rienitz},\
  and\ \citenamefont {Schiel}}]{Zakel:2011}%
  \BibitemOpen
  \bibfield  {author} {\bibinfo {author} {\bibfnamefont {S.}~\bibnamefont
  {Zakel}}, \bibinfo {author} {\bibfnamefont {S.}~\bibnamefont {Wundrack}},
  \bibinfo {author} {\bibfnamefont {H.}~\bibnamefont {Niemann}}, \bibinfo
  {author} {\bibfnamefont {O.}~\bibnamefont {Rienitz}}, \ and\ \bibinfo
  {author} {\bibfnamefont {D.}~\bibnamefont {Schiel}},\ }\href@noop {}
  {\bibfield  {journal} {\bibinfo  {journal} {Metrologia}\ }\textbf {\bibinfo
  {volume} {48}},\ \bibinfo {pages} {S14} (\bibinfo {year} {2011})}\BibitemShut
  {NoStop}%
\bibitem [{\citenamefont {Busch}\ \emph {et~al.}(2011)\citenamefont {Busch},
  \citenamefont {Azuma}, \citenamefont {Bettin}, \citenamefont {Cibik},
  \citenamefont {Fuchs}, \citenamefont {Fujii}, \citenamefont {Krumrey},
  \citenamefont {Kuetgens}, \citenamefont {Kuramoto},\ and\ \citenamefont
  {Mizushima}}]{Busch:2011}%
  \BibitemOpen
  \bibfield  {author} {\bibinfo {author} {\bibfnamefont {I.}~\bibnamefont
  {Busch}}, \bibinfo {author} {\bibfnamefont {Y.}~\bibnamefont {Azuma}},
  \bibinfo {author} {\bibfnamefont {H.}~\bibnamefont {Bettin}}, \bibinfo
  {author} {\bibfnamefont {L.}~\bibnamefont {Cibik}}, \bibinfo {author}
  {\bibfnamefont {P.}~\bibnamefont {Fuchs}}, \bibinfo {author} {\bibfnamefont
  {K.}~\bibnamefont {Fujii}}, \bibinfo {author} {\bibfnamefont
  {M.}~\bibnamefont {Krumrey}}, \bibinfo {author} {\bibfnamefont
  {U.}~\bibnamefont {Kuetgens}}, \bibinfo {author} {\bibfnamefont
  {N.}~\bibnamefont {Kuramoto}}, \ and\ \bibinfo {author} {\bibfnamefont
  {S.}~\bibnamefont {Mizushima}},\ }\href@noop {} {\bibfield  {journal}
  {\bibinfo  {journal} {Metrologia}\ }\textbf {\bibinfo {volume} {48}},\
  \bibinfo {pages} {S62} (\bibinfo {year} {2011})}\BibitemShut {NoStop}%
\bibitem [{\citenamefont {Ferroglio}, \citenamefont {Mana},\ and\ \citenamefont
  {Massa}(2008)}]{Ferroglio:2008}%
  \BibitemOpen
  \bibfield  {author} {\bibinfo {author} {\bibfnamefont {L.}~\bibnamefont
  {Ferroglio}}, \bibinfo {author} {\bibfnamefont {G.}~\bibnamefont {Mana}}, \
  and\ \bibinfo {author} {\bibfnamefont {E.}~\bibnamefont {Massa}},\
  }\href@noop {} {\bibfield  {journal} {\bibinfo  {journal} {Opt.\ Express}\
  }\textbf {\bibinfo {volume} {16}},\ \bibinfo {pages} {16877} (\bibinfo {year}
  {2008})}\BibitemShut {NoStop}%
\bibitem [{\citenamefont {Bergamin}, \citenamefont {Cavagnero},\ and\
  \citenamefont {Mana}(1991)}]{Mana:1991}%
  \BibitemOpen
  \bibfield  {author} {\bibinfo {author} {\bibfnamefont {A.}~\bibnamefont
  {Bergamin}}, \bibinfo {author} {\bibfnamefont {G.}~\bibnamefont {Cavagnero}},
  \ and\ \bibinfo {author} {\bibfnamefont {G.}~\bibnamefont {Mana}},\
  }\href@noop {} {\bibfield  {journal} {\bibinfo  {journal} {Meas.\ Sci.\
  Technol.}\ }\textbf {\bibinfo {volume} {2}},\ \bibinfo {pages} {725}
  (\bibinfo {year} {1991})}\BibitemShut {NoStop}%
\bibitem [{\citenamefont {Paetzelt}\ \emph {et~al.}(2013)\citenamefont
  {Paetzelt}, \citenamefont {Arnold}, \citenamefont {B\"ohm}, \citenamefont
  {Pietag},\ and\ \citenamefont {Schindler}}]{Paetzelt:2013}%
  \BibitemOpen
  \bibfield  {author} {\bibinfo {author} {\bibfnamefont {H.}~\bibnamefont
  {Paetzelt}}, \bibinfo {author} {\bibfnamefont {T.}~\bibnamefont {Arnold}},
  \bibinfo {author} {\bibfnamefont {G.}~\bibnamefont {B\"ohm}}, \bibinfo
  {author} {\bibfnamefont {F.}~\bibnamefont {Pietag}}, \ and\ \bibinfo {author}
  {\bibfnamefont {A.}~\bibnamefont {Schindler}},\ }\href@noop {} {\bibfield
  {journal} {\bibinfo  {journal} {Plasma Processes and Polymers}\ }\textbf
  {\bibinfo {volume} {10}},\ \bibinfo {pages} {416} (\bibinfo {year}
  {2013})}\BibitemShut {NoStop}%
\bibitem [{\citenamefont {Paetzelt}, \citenamefont {B\"ohm},\ and\
  \citenamefont {Arnold}(2014)}]{Paetzelt:2014}%
  \BibitemOpen
  \bibfield  {author} {\bibinfo {author} {\bibfnamefont {H.}~\bibnamefont
  {Paetzelt}}, \bibinfo {author} {\bibfnamefont {G.}~\bibnamefont {B\"ohm}}, \
  and\ \bibinfo {author} {\bibfnamefont {T.}~\bibnamefont {Arnold}},\
  }\href@noop {} {\bibfield  {journal} {\bibinfo  {journal} {Plasma Sources
  Sci.\ Technol.}\ ,\ \bibinfo {pages} {submitted}} (\bibinfo {year}
  {2014})}\BibitemShut {NoStop}%
\bibitem [{\citenamefont {Bergamin}\ \emph
  {et~al.}(1999{\natexlab{a}})\citenamefont {Bergamin}, \citenamefont
  {Cavagnero}, \citenamefont {Cordiali},\ and\ \citenamefont
  {Mana}}]{Mana:1999a}%
  \BibitemOpen
  \bibfield  {author} {\bibinfo {author} {\bibfnamefont {A.}~\bibnamefont
  {Bergamin}}, \bibinfo {author} {\bibfnamefont {G.}~\bibnamefont {Cavagnero}},
  \bibinfo {author} {\bibfnamefont {L.}~\bibnamefont {Cordiali}}, \ and\
  \bibinfo {author} {\bibfnamefont {G.}~\bibnamefont {Mana}},\ }\href@noop {}
  {\bibfield  {journal} {\bibinfo  {journal} {Eur.\ Phys.\ J.\ D}\ }\textbf
  {\bibinfo {volume} {5}},\ \bibinfo {pages} {433} (\bibinfo {year}
  {1999}{\natexlab{a}})}\BibitemShut {NoStop}%
\bibitem [{\citenamefont {Cavagnero}, \citenamefont {Mana},\ and\ \citenamefont
  {Massa}(2006)}]{Mana:2006}%
  \BibitemOpen
  \bibfield  {author} {\bibinfo {author} {\bibfnamefont {G.}~\bibnamefont
  {Cavagnero}}, \bibinfo {author} {\bibfnamefont {G.}~\bibnamefont {Mana}}, \
  and\ \bibinfo {author} {\bibfnamefont {E.}~\bibnamefont {Massa}},\
  }\href@noop {} {\bibfield  {journal} {\bibinfo  {journal} {J.\ Opt.\ Soc.\
  Am.\ A}\ }\textbf {\bibinfo {volume} {23}},\ \bibinfo {pages} {1951}
  (\bibinfo {year} {2006})}\BibitemShut {NoStop}%
\bibitem [{\citenamefont {Mana}, \citenamefont {Massa},\ and\ \citenamefont
  {Rovera}(2001)}]{Mana:2001}%
  \BibitemOpen
  \bibfield  {author} {\bibinfo {author} {\bibfnamefont {G.}~\bibnamefont
  {Mana}}, \bibinfo {author} {\bibfnamefont {E.}~\bibnamefont {Massa}}, \ and\
  \bibinfo {author} {\bibfnamefont {A.}~\bibnamefont {Rovera}},\ }\href@noop {}
  {\bibfield  {journal} {\bibinfo  {journal} {Appl.\ Opt.}\ }\textbf {\bibinfo
  {volume} {40}},\ \bibinfo {pages} {1378} (\bibinfo {year}
  {2001})}\BibitemShut {NoStop}%
\bibitem [{\citenamefont {Bergamin}\ \emph
  {et~al.}(1999{\natexlab{b}})\citenamefont {Bergamin}, \citenamefont
  {Cavagnero}, \citenamefont {Mana}, \citenamefont {Massa},\ and\ \citenamefont
  {Zosi}}]{Mana:1999b}%
  \BibitemOpen
  \bibfield  {author} {\bibinfo {author} {\bibfnamefont {A.}~\bibnamefont
  {Bergamin}}, \bibinfo {author} {\bibfnamefont {G.}~\bibnamefont {Cavagnero}},
  \bibinfo {author} {\bibfnamefont {G.}~\bibnamefont {Mana}}, \bibinfo {author}
  {\bibfnamefont {E.}~\bibnamefont {Massa}}, \ and\ \bibinfo {author}
  {\bibfnamefont {G.}~\bibnamefont {Zosi}},\ }\href@noop {} {\bibfield
  {journal} {\bibinfo  {journal} {Meas.\ Sci.\ Technol.}\ }\textbf {\bibinfo
  {volume} {10}},\ \bibinfo {pages} {549} (\bibinfo {year}
  {1999}{\natexlab{b}})}\BibitemShut {NoStop}%
\bibitem [{\citenamefont {Massa}\ and\ \citenamefont
  {Mana}(2013)}]{Massa:2013}%
  \BibitemOpen
  \bibfield  {author} {\bibinfo {author} {\bibfnamefont {E.}~\bibnamefont
  {Massa}}\ and\ \bibinfo {author} {\bibfnamefont {G.}~\bibnamefont {Mana}},\
  }\href@noop {} {\bibfield  {journal} {\bibinfo  {journal} {Meas.\ Sci.\
  Technol.}\ }\textbf {\bibinfo {volume} {24}},\ \bibinfo {pages} {107001}
  (\bibinfo {year} {2013})}\BibitemShut {NoStop}%
\bibitem [{\citenamefont {White}\ \emph {et~al.}(2008)\citenamefont {White},
  \citenamefont {Ballico}, \citenamefont {Chimenti}, \citenamefont {Duris},
  \citenamefont {{Fi\-li\-pe}}, \citenamefont {{I\-va\-nova}}, \citenamefont
  {Dogan}, \citenamefont {Mendez-Lango}, \citenamefont {Meyer}, \citenamefont
  {Pavese}, \citenamefont {Peruzzi}, \citenamefont {Renaot}, \citenamefont
  {Rudtsch},\ and\ \citenamefont {Yamazawa}}]{CCT/08-19/rev}%
  \BibitemOpen
  \bibfield  {author} {\bibinfo {author} {\bibfnamefont {D.}~\bibnamefont
  {White}}, \bibinfo {author} {\bibfnamefont {M.}~\bibnamefont {Ballico}},
  \bibinfo {author} {\bibfnamefont {V.}~\bibnamefont {Chimenti}}, \bibinfo
  {author} {\bibfnamefont {S.}~\bibnamefont {Duris}}, \bibinfo {author}
  {\bibfnamefont {E.}~\bibnamefont {{Fi\-li\-pe}}}, \bibinfo {author}
  {\bibfnamefont {A.}~\bibnamefont {{I\-va\-nova}}}, \bibinfo {author}
  {\bibfnamefont {A.~K.}\ \bibnamefont {Dogan}}, \bibinfo {author}
  {\bibfnamefont {E.}~\bibnamefont {Mendez-Lango}}, \bibinfo {author}
  {\bibfnamefont {C.}~\bibnamefont {Meyer}}, \bibinfo {author} {\bibfnamefont
  {F.}~\bibnamefont {Pavese}}, \bibinfo {author} {\bibfnamefont
  {A.}~\bibnamefont {Peruzzi}}, \bibinfo {author} {\bibfnamefont
  {E.}~\bibnamefont {Renaot}}, \bibinfo {author} {\bibfnamefont
  {S.}~\bibnamefont {Rudtsch}}, \ and\ \bibinfo {author} {\bibfnamefont
  {K.}~\bibnamefont {Yamazawa}},\ }\href@noop {} {\enquote {\bibinfo {title}
  {Uncertainties in the realisation of the {SPRT} subranges of the {ITS-90}},}\
  }\bibinfo {type} {Report of the CCT-WG3 on Uncertainties in Contact
  Thermometry}\ \bibinfo {number} {CCT/08-19/rev}\ (\bibinfo  {institution}
  {Bureau International des Poids et Mesures},\ \bibinfo {address} {S\`evres},\
  \bibinfo {year} {2008})\BibitemShut {NoStop}%
\bibitem [{\citenamefont {Bartl}\ \emph {et~al.}(2009)\citenamefont {Bartl},
  \citenamefont {Nicolaus}, \citenamefont {Kessler}, \citenamefont
  {Sch\"odel},\ and\ \citenamefont {Becker}}]{Bartl:2009}%
  \BibitemOpen
  \bibfield  {author} {\bibinfo {author} {\bibfnamefont {G.}~\bibnamefont
  {Bartl}}, \bibinfo {author} {\bibfnamefont {A.}~\bibnamefont {Nicolaus}},
  \bibinfo {author} {\bibfnamefont {E.}~\bibnamefont {Kessler}}, \bibinfo
  {author} {\bibfnamefont {R.}~\bibnamefont {Sch\"odel}}, \ and\ \bibinfo
  {author} {\bibfnamefont {P.}~\bibnamefont {Becker}},\ }\href@noop {}
  {\bibfield  {journal} {\bibinfo  {journal} {Metrologia}\ }\textbf {\bibinfo
  {volume} {46}},\ \bibinfo {pages} {416} (\bibinfo {year} {2009})}\BibitemShut
  {NoStop}%
\bibitem [{\citenamefont {Mana}, \citenamefont {Massa},\ and\ \citenamefont
  {Ferroglio}(2009)}]{Mana:2009}%
  \BibitemOpen
  \bibfield  {author} {\bibinfo {author} {\bibfnamefont {G.}~\bibnamefont
  {Mana}}, \bibinfo {author} {\bibfnamefont {E.}~\bibnamefont {Massa}}, \ and\
  \bibinfo {author} {\bibfnamefont {L.}~\bibnamefont {Ferroglio}},\ }\href@noop
  {} {\bibfield  {journal} {\bibinfo  {journal} {Opt.\ Express}\ }\textbf
  {\bibinfo {volume} {17}},\ \bibinfo {pages} {11172} (\bibinfo {year}
  {2009})}\BibitemShut {NoStop}%
\bibitem [{\citenamefont {Ferroglio}, \citenamefont {Mana},\ and\ \citenamefont
  {Massa}(2011)}]{Ferroglio:2011}%
  \BibitemOpen
  \bibfield  {author} {\bibinfo {author} {\bibfnamefont {L.}~\bibnamefont
  {Ferroglio}}, \bibinfo {author} {\bibfnamefont {G.}~\bibnamefont {Mana}}, \
  and\ \bibinfo {author} {\bibfnamefont {E.}~\bibnamefont {Massa}},\
  }\href@noop {} {\bibfield  {journal} {\bibinfo  {journal} {Metrologia}\
  }\textbf {\bibinfo {volume} {48}},\ \bibinfo {pages} {S50} (\bibinfo {year}
  {2011})}\BibitemShut {NoStop}%
\bibitem [{\citenamefont {Mana}\ and\ \citenamefont
  {Vittone}(1997{\natexlab{a}})}]{Mana:1997a}%
  \BibitemOpen
  \bibfield  {author} {\bibinfo {author} {\bibfnamefont {G.}~\bibnamefont
  {Mana}}\ and\ \bibinfo {author} {\bibfnamefont {E.}~\bibnamefont {Vittone}},\
  }\href@noop {} {\bibfield  {journal} {\bibinfo  {journal} {Z.\ Phys.\ B}\
  }\textbf {\bibinfo {volume} {102}},\ \bibinfo {pages} {189} (\bibinfo {year}
  {1997}{\natexlab{a}})}\BibitemShut {NoStop}%
\bibitem [{\citenamefont {Mana}\ and\ \citenamefont
  {Vittone}(1997{\natexlab{b}})}]{Mana:1997b}%
  \BibitemOpen
  \bibfield  {author} {\bibinfo {author} {\bibfnamefont {G.}~\bibnamefont
  {Mana}}\ and\ \bibinfo {author} {\bibfnamefont {E.}~\bibnamefont {Vittone}},\
  }\href@noop {} {\bibfield  {journal} {\bibinfo  {journal} {Z.\ Phys.\ B}\
  }\textbf {\bibinfo {volume} {102}},\ \bibinfo {pages} {197} (\bibinfo {year}
  {1997}{\natexlab{b}})}\BibitemShut {NoStop}%
\bibitem [{\citenamefont {Quagliotti}\ \emph {et~al.}(2013)\citenamefont
  {Quagliotti}, \citenamefont {Mana}, \citenamefont {Massa}, \citenamefont
  {Sasso},\ and\ \citenamefont {Kuetgens}}]{Quagliotti:2013}%
  \BibitemOpen
  \bibfield  {author} {\bibinfo {author} {\bibfnamefont {D.}~\bibnamefont
  {Quagliotti}}, \bibinfo {author} {\bibfnamefont {G.}~\bibnamefont {Mana}},
  \bibinfo {author} {\bibfnamefont {E.}~\bibnamefont {Massa}}, \bibinfo
  {author} {\bibfnamefont {C.}~\bibnamefont {Sasso}}, \ and\ \bibinfo {author}
  {\bibfnamefont {U.}~\bibnamefont {Kuetgens}},\ }\href@noop {} {\bibfield
  {journal} {\bibinfo  {journal} {Metrologia}\ }\textbf {\bibinfo {volume}
  {50}},\ \bibinfo {pages} {243} (\bibinfo {year} {2013})}\BibitemShut
  {NoStop}%
\bibitem [{\citenamefont {Melis}, \citenamefont {Colombo},\ and\ \citenamefont
  {Mana}(2014)}]{Melis:2014}%
  \BibitemOpen
  \bibfield  {author} {\bibinfo {author} {\bibfnamefont {C.}~\bibnamefont
  {Melis}}, \bibinfo {author} {\bibfnamefont {L.}~\bibnamefont {Colombo}}, \
  and\ \bibinfo {author} {\bibfnamefont {G.}~\bibnamefont {Mana}},\ }\href@noop
  {} {\bibfield  {journal} {\bibinfo  {journal} {Metrologia}\ }\textbf
  {\bibinfo {volume} {52}},\ \bibinfo {pages} {214} (\bibinfo {year}
  {2014})}\BibitemShut {NoStop}%
\end{thebibliography}
\providecommand{\noopsort}[1]{}\providecommand{\singleletter}[1]{#1}%

\end{document}